\documentclass[twocolumn]{aastex63}
\usepackage[utf8]{inputenc}

\usepackage{natbib}
\usepackage{graphicx}
\usepackage{amsmath}
\usepackage{booktabs}
\usepackage{longtable}
\usepackage{xspace}
\usepackage{hyperref}
\usepackage[T1]{fontenc}
\usepackage{lipsum}
\usepackage{comment}
\usepackage{xcolor}
\usepackage{rotating}

\usepackage{IEEEtrantools} 
\usepackage{amsmath}

\graphicspath{{Figures/}} 

\newcommand{\ms}{\ensuremath{\rm m\,s^{-1}}}

\newcommand{\gcmc}{\ensuremath{\rm g\,cm^{-3}}}
\newcommand{\gcc}{\gcmc}

\newcommand{\teff}{\ensuremath{T_{\rm eff}}}
\newcommand{\logg}{\ensuremath{\log{g}}}
\newcommand{\vsini}{\ensuremath{v \sin{i}}}
\newcommand{\feh}{\ensuremath{\mathrm{[Fe/H]}}}

\newcommand{\rsun}{\ensuremath{R_\sun}}
\newcommand{\msun}{\ensuremath{M_\sun}}
\newcommand{\lsun}{\ensuremath{L_\sun}}

\newcommand{\rstar}{\ensuremath{R_\star}}
\newcommand{\mstar}{\ensuremath{M_\star}}

\newcommand{\lstar}{\ensuremath{L_\star}}

\newcommand{\rhostar}{\ensuremath{\rho_\star}}

\newcommand{\rpl}{\ensuremath{R_{\rm p}}}
\newcommand{\mpl}{\ensuremath{M_{\rm p}}}

\newcommand{\rhopl}{\ensuremath{\rho_{\rm p}}}

\newcommand{\mjup}{\ensuremath{M_{\rm J}}}

\newcommand{\rearth}{\ensuremath{R_\earth}}
\newcommand{\mearth}{\ensuremath{M_\earth}}

\newcommand{\mplsini}{\ensuremath{\mpl\sin{i}}}

\newcommand{\Kepler}{\textit{Kepler}}

\newcommand{\K}{\ensuremath{K}}

\newcommand{\nrvs}{60}
\newcommand{\lastobs}{October}

\newcommand{\rhosun}{\ensuremath{\rho_{\odot}}}

\newcommand{\afeh}{\mbox{$\rm{[\alpha/Fe]}$}}
\newcommand{\mh}{\mbox{$\rm{[M/H]}$}}

\newcommand{\target}{TOI-561}
\newcommand{\tess}{{\it TESS}}
\newcommand{\gaia}{\mbox{\textit{Gaia}}}

\newcommand{\radstar}{\mbox{$0.832\pm0.019$}}
\newcommand{\massstar}{\mbox{$0.805\pm0.030$}}
\newcommand{\loggseis}{\mbox{$4.500\pm0.030$}}
\newcommand{\denstar}{\mbox{$1.38\pm0.11$}}

\newcommand{\teffstar}{\mbox{$5326 \pm 64$}}
\newcommand{\fehstar}{\mbox{$-0.41 \pm 0.05$}}
\newcommand{\afestar}{\mbox{$+0.23 \pm 0.05$}}

\newcommand{\rb}{\ensuremath{1.45\pm 0.11}}
\newcommand{\rc}{\ensuremath{2.90\pm 0.13}}
\newcommand{\rd}{\ensuremath{2.32\pm 0.16}}

\newcommand{\KbA}{\ensuremath{3.1\pm 0.8}}
\newcommand{\KcA}{\ensuremath{2.2\pm 0.8}}
\newcommand{\KdA}{\ensuremath{0.3\pm 0.8}}

\newcommand{\KbB}{\ensuremath{3.1 \pm 0.8}}
\newcommand{\KcB}{\ensuremath{2.4 \pm 0.8}}
\newcommand{\KdB}{\ensuremath{0.9 \pm 0.6}}

\newcommand{\KbC}{\ensuremath{3.2\pm 0.8}}
\newcommand{\KcC}{\ensuremath{2.1\pm0.8}}
\newcommand{\KdC}{\ensuremath{1.0^{+0.7}_{-0.6}}}

\newcommand{\gdot}{\ensuremath{0.009\pm0.004 \ms}}

\newcommand{\mb}{\ensuremath{3.2 \pm 0.8}}
\newcommand{\mc}{\ensuremath{7.0\pm 2.3}}
\newcommand{\md}{\ensuremath{3.0^{+2.4}_{-1.9}}}

\newcommand{\rhob}{\ensuremath{5.5^{+2.0}_{-1.6}}}
\newcommand{\rhoc}{\ensuremath{1.6\pm0.6}}
\newcommand{\rhod}{\ensuremath{1.3^{+1.1}_{-0.8}}}

\submitjournal{The Astronomical Journal}
\received{Sept. 04, 2020}
\revised {Dec. 04, 2020}
\accepted{Dec. 14, 2020}

\begin{document}
\title{The {\it TESS}-Keck Survey II: An Ultra-Short Period Rocky Planet and its Siblings Transiting the Galactic Thick-Disk Star TOI-561}
\suppressAffiliations

\author[0000-0002-3725-3058]{Lauren M. Weiss} 
\affiliation{Institute for Astronomy, University of Hawai`i, 2680 Woodlawn Drive, Honolulu, HI 96822, USA}

\author[0000-0002-8958-0683]{Fei Dai} 
\affiliation{Division of Geological and Planetary Sciences,
1200 E California Blvd, Pasadena, CA, 91125, USA}

\author[0000-0001-8832-4488]{Daniel Huber} 
\affiliation{Institute for Astronomy, University of Hawai`i, 2680 Woodlawn Drive, Honolulu, HI 96822, USA}

\author[0000-0002-9873-1471]{John M. Brewer} 
\affiliation{Department of Physics and Astronomy, San Francisco State University, 1600 Holloway Ave, San Francisco, CA 94132, USA}

\author[0000-0001-6588-9574]{Karen A.\ Collins} 
\affiliation{Center for Astrophysics \textbar \ Harvard \& Smithsonian, 60 Garden Street, Cambridge, MA 02138, USA}

\author[0000-0002-5741-3047]{David R. Ciardi} 
\affiliation{NASA Exoplanet Science Institute, Caltech/IPAC, Pasadena, CA 91125}

\author[0000-0003-0593-1560]{Elisabeth C. Matthews} 
\affiliation{Observatorie de l\'{U}niversit\'{e} de Gen\`{e}ve, 51 Chemin des Maillettes, 1290 Versoix, Switzerland}

\author[0000-0002-0619-7639]{Carl Ziegler} 
\affiliation{Dunlap Institute for Astronomy and Astrophysics, University of Toronto, 50 St. George Street, Toronto, Ontario M5S 3H4, Canada}

\author[0000-0002-2532-2853]{Steve B. Howell} 
\affiliation{NASA Ames Research Center, Moffett Field, CA 94035, USA}

\author[0000-0002-7030-9519]{Natalie M. Batalha}
\affiliation{Department of Astronomy and Astrophysics, University of California, Santa Cruz, CA 95060, USA}

\author{Ian J. M. Crossfield}
\affiliation{Department of Physics \& Astronomy, University of Kansas, 1082 Malott,1251 Wescoe Hall Dr., Lawrence, KS 66045, USA}

\author[0000-0001-8189-0233]{Courtney Dressing}
\affiliation{501 Campbell Hall, University of California at Berkeley, Berkeley, CA 94720, USA}

\author[0000-0003-3504-5316]{Benjamin Fulton}
\affiliation{NASA Exoplanet Science Institute/Caltech-IPAC, MC 314-6, 1200 E California Blvd, Pasadena, CA 91125, USA}

\author[0000-0001-8638-0320]{Andrew W. Howard}
\affiliation{Department of Astronomy, California Institute of Technology, Pasadena, CA 91125, USA}

\author[0000-0002-0531-1073]{Howard Isaacson}
\affiliation{{Department of Astronomy,  University of California Berkeley, Berkeley CA 94720, USA}}
\affiliation{Centre for Astrophysics, University of Southern Queensland, Toowoomba, QLD, Australia}

\author[0000-0002-7084-0529]{Stephen R. Kane}
\affiliation{Department of Earth and Planetary Sciences, University of California, Riverside, CA 92521, USA}

\author[0000-0003-0967-2893]{Erik A Petigura}
\affiliation{Department of Physics \& Astronomy, University of California Los Angeles, Los Angeles, CA 90095, USA}

\author[0000-0003-0149-9678]{Paul Robertson}
\affiliation{Department of Physics \& Astronomy, University of California Irvine, Irvine, CA 92697, USA}

\author[0000-0001-8127-5775]{Arpita Roy}
\affiliation{Department of Astronomy, California Institute of Technology, Pasadena, CA 91125, USA}

\author[0000-0003-3856-3143]{Ryan A. Rubenzahl}
\altaffiliation{NSF Graduate Research Fellow}
\affiliation{Department of Astronomy, California Institute of Technology, Pasadena, CA 91125, USA}

\author[0000-0002-6778-7552]{Joseph D. Twicken}
\affiliation{SETI Institute, Mountain View, CA  94043, USA}
\affiliation{NASA Ames Research Center, Moffett Field, CA  94035, USA}

\author[0000-0002-9879-3904]{Zachary R. Claytor}
\affiliation{Institute for Astronomy, University of Hawai`i, 2680 Woodlawn Drive, Honolulu, HI 96822, USA}

\author[0000-0002-3481-9052]{Keivan G. Stassun}
\affiliation{Vanderbilt University, Department of Physics \& Astronomy, 6301 Stevenson Center Ln., Nashville, TN 37235, USA }

\author[0000-0003-2562-9043]{Mason G. MacDougall}
\affiliation{Department of Physics \& Astronomy, University of California Los Angeles, Los Angeles, CA 90095, USA}

\author[0000-0003-1125-2564]{Ashley Chontos}
\altaffiliation{NSF Graduate Research Fellow}
\affiliation{Institute for Astronomy, University of Hawai`i, 2680 Woodlawn Drive, Honolulu, HI 96822, USA}

\author[0000-0002-8965-3969]{Steven Giacalone}
\affil{Department of Astronomy, University of California Berkeley, Berkeley, CA 94720, USA}

\author[0000-0002-4297-5506]{Paul A. Dalba}
\altaffiliation{NSF Astronomy and Astrophysics Postdoctoral Fellow}
\affiliation{Department of Earth and Planetary Sciences, University of California, Riverside, CA 92521, USA}

\author[0000-0003-4603-556X]{Teo Mocnik}
\affiliation{Gemini Observatory, Northern Operations Center, 670 N. A'ohoku Place, Hilo, HI 96720, USA}

\author[0000-0002-0139-4756]{Michelle L. Hill}
\affiliation{Department of Earth and Planetary Sciences, University of California, Riverside, CA 92521, USA}

\author[0000-0001-7708-2364]{Corey Beard}
\affiliation{Department of Physics \& Astronomy, University of California Irvine, Irvine, CA 92697, USA}

\author[0000-0001-8898-8284]{Joseph M. Akana Murphy}
\altaffiliation{NSF Graduate Research Fellow}
\affiliation{Department of Astronomy and Astrophysics, University of California, Santa Cruz, CA 95064, USA}

\author[0000-0001-8391-5182]{Lee J.\ Rosenthal}
\affiliation{Department of Astronomy, California Institute of Technology, Pasadena, CA 91125, USA}

\author[0000-0003-0012-9093]{Aida Behmard}
\altaffiliation{NSF Graduate Research Fellow}
\affiliation{Division of Geological and Planetary Science, California Institute of Technology, Pasadena, CA 91125, USA}

\author[0000-0002-4290-6826]{Judah Van Zandt}
\affiliation{Department of Physics \& Astronomy, University of California Los Angeles, Los Angeles, CA 90095, USA}

\author[0000-0001-8342-7736]{Jack Lubin}
\affiliation{Department of Physics \& Astronomy, University of California Irvine, Irvine, CA 92697, USA}

\author[0000-0002-6115-4359]{Molly R. Kosiarek}
\affiliation{Department of Astronomy and Astrophysics, University of California, Santa Cruz, CA 95064, USA}
\affiliation{NSF Graduate Research Fellow} 

\author[0000-0003-2527-1598]{Michael B. Lund}
\affiliation{NASA Exoplanet Science Institute, Caltech/IPAC, Pasadena, CA 91125, USA}

\author[0000-0002-8035-4778]{Jessie L. Christiansen}
\affiliation{NASA Exoplanet Science Institute, Caltech/IPAC, Pasadena, CA 91125, USA}

\author[0000-0001-7233-7508]{Rachel A. Matson}
\affiliation{US Naval Observatory, 3450 Massachusetts Ave. NW., Washington, DC 20392, USA}

\author{Charles A. Beichman}
\affiliation{NASA Exoplanet Science Institute, Caltech/IPAC, Pasadena, CA 91125}

\author[0000-0001-5347-7062]{Joshua E. Schlieder}
\affiliation{Exoplanets and Stellar Astrophysics Laboratory, Mail Code 667, NASA Goddard Space Flight Center, Greenbelt, MD 20771, USA}

\author[0000-0002-9329-2190]{Erica J. Gonzales}
\affiliation{University of California, Santa Cruz, 1156 High St., Santa Cruz, CA, 95064, USA}
\affiliation{National Science Foundation Graduate Research Fellow}

\author[0000-0001-7124-4094]{C\'{e}sar Brice\~{n}o}
\affiliation{CCerro Tololo Inter-American Observatory, Casilla 603, La Serena 1700000, Chile} 

\author{Nicholas Law}
\affiliation{Department of Physics and Astronomy, The University of North Carolina at Chapel Hill, Chapel Hill, NC 27599-3255, USA}

\author[0000-0003-3654-1602]{Andrew W. Mann}
\affiliation{Department of Physics and Astronomy, The University of North Carolina at Chapel Hill, Chapel Hill, NC 27599-3255, USA}


\author[0000-0003-2781-3207]{Kevin I.\ Collins}
\affiliation{George Mason University, 4400 University Drive, Fairfax, VA, 22030 USA}
\author[0000-0002-5674-2404]{Phil Evans}
\affiliation{El Sauce Observatory, Coquimbo Province, Chile}

\author[0000-0002-4909-5763]{Akihiko  Fukui}
\affiliation{Department of Earth and Planetary Science, Graduate School of Science, The University of Tokyo, 7-3-1 Hongo, Bunkyo-ku, Tokyo 113-0033, Japan}
\affiliation{Instituto de Astrof\'isica de Canarias, V\'ia L\'actea s/n, E-38205 La Laguna, Tenerife, Spain}

\author[0000-0002-4625-7333]{Eric L. N. Jensen}
\affiliation{Dept.\ of Physics \& Astronomy, Swarthmore College, Swarthmore PA 19081, USA}

\author{Felipe Murgas}
\affiliation{Instituto de Astrof\'\i sica de Canarias (IAC), 38205 La Laguna, Tenerife, Spain}
\affiliation{Departamento de Astrof\'\i sica, Universidad de La Laguna (ULL), 38206, La Laguna, Tenerife, Spain}

\author[0000-0001-8511-2981]{Norio Narita}
\affiliation{Komaba Institute for Science, The University of Tokyo, 3-8-1 Komaba, Meguro, Tokyo 153-8902, Japan}
\affiliation{JST, PRESTO, 3-8-1 Komaba, Meguro, Tokyo 153-8902, Japan}
\affiliation{Astrobiology Center, 2-21-1 Osawa, Mitaka, Tokyo 181-8588, Japan}
\affiliation{Instituto de Astrof\'{i}sica de Canarias (IAC), 38205 La Laguna, Tenerife, Spain}

\author[0000-0003-0987-1593]{Enric Palle}
\affiliation{Instituto de Astrof\'\i sica de Canarias (IAC), 38205 La Laguna, Tenerife, Spain}
\affiliation{Departamento de Astrof\'\i sica, Universidad de La Laguna (ULL), 38206, La Laguna, Tenerife, Spain}

\author[0000-0001-5519-1391]{Hannu Parviainen}
\affiliation{Instituto de Astrof\'\i sica de Canarias (IAC), 38205 La Laguna, Tenerife, Spain}
\affiliation{Departamento de Astrof\'\i sica, Universidad de La Laguna (ULL), 38206, La Laguna, Tenerife, Spain}

\author[0000-0001-8227-1020]{Richard P. Schwarz}
\affiliation{Patashnick Voorheesville Observatory, Voorheesville, NY 12186, USA}

\author[0000-0001-5603-6895]{Thiam-Guan Tan}
\affiliation{Perth Exoplanet Survey Telescope, Perth, Western Australia}

\author[0000-0002-1860-7842]{Jack S.~Acton}
\affiliation{School of Physics and Astronomy,\\
University of Leicester, University Road, Leicester, LE1 7RH, UK}

\author[0000-0001-7904-4441]{Edward M.~Bryant}
\affiliation{Dept.\ of Physics,\\
University of Warwick, Gibbet Hill Road,\\
Coventry CV4 7AL, UK}

\author[0000-0003-0061-5446]{Alexander Chaushev}
\affiliation{Center for Astronomy and Astrophysics,\\
TU Berlin, Hardenbergstr. 36, D-10623 Berlin, Germany}

\author[0000-0003-4096-0594]{Philipp Eigm\"uller}
\affiliation{Institute of Planetary Research,German Aerospace Center,\\
Rutherfordstrasse 2, 12489 Berlin, Germany}

\author[0000-0002-4259-0155]{Sam Gill}
\affiliation{Dept.\ of Physics,\\
University of Warwick, Gibbet Hill Road,\\
Coventry CV4 7AL, UK}


\author[0000-0002-4715-9460]{Jon Jenkins}
\affiliation{NASA Ames Research Center, Moffett Field, CA 94035, USA}

\author[0000-0003-2058-6662]{George Ricker}
\affiliation{Kavli Institute for Astrophysics and Space Research, Massachusetts Institute of Technology, Cambridge, MA 02139, USA}

\author[0000-0002-6892-6948]{Sara Seager}
\affiliation{Department of Physics and Kavli Institute for Astrophysics and Space Research, Massachusetts Institute of Technology, Cambridge, MA 02139, USA}
\affiliation{Department of Earth, Atmospheric and Planetary Sciences, Massachusetts Institute of Technology, Cambridge, MA 02139, USA}
\affiliation{Department of Aeronautics and Astronautics, MIT, 77 Massachusetts Avenue, Cambridge, MA 02139, USA}


\author[0000-0002-4265-047X]{Joshua N.\ Winn}
\affiliation{Department of Astrophysical Sciences, Princeton University, 4 Ivy Lane, Princeton, NJ 08544, USA}

\newcommand{\TESS}{\textit{TESS}}

\begin{abstract}
We report the discovery of TOI-561, a multi-planet system in the galactic thick disk that contains a rocky, ultra-short period planet (USP).  This bright ($V=10.2$) star hosts three small transiting planets identified in photometry from the NASA \textit{TESS} mission: TOI-561 b (TOI-561.02, P=0.44 days, \rpl=\rb\,\rearth), c (TOI-561.01, P=10.8 days, \rpl= \rc\,\rearth), and d (TOI-561.03, P=16.3 days, \rpl=\rd\,\rearth).  The star is chemically (\feh = \fehstar, \afeh=\afestar) and kinematically consistent with the galactic thick disk population, making TOI-561 one of the oldest ($10\pm3\,$Gyr) and most metal-poor planetary systems discovered yet.  We dynamically confirm planets b and c with radial velocities from the W. M. Keck Observatory High Resolution Echelle Spectrometer.  Planet b has a mass and density of \mb\,\mearth\ and \rhob\,\gcc, consistent with a rocky composition.  Its lower-than-average density is consistent with an iron-poor composition, although an Earth-like iron-to-silicates ratio is not ruled out.  Planet c is \mc\,\mearth\ and \rhoc\,\gcc, consistent with an interior rocky core overlaid with a low-mass volatile envelope.  Several attributes of the photometry for planet d (which we did not detect dynamically) complicate the analysis, but we vet the planet with high-contrast imaging, ground-based photometric follow-up and radial velocities.  TOI-561 b is the first rocky world around a galactic thick-disk star confirmed with radial velocities and one of the best rocky planets for thermal emission studies.  
\end{abstract}


\section{Introduction \label{sec:intro}} 
The NASA \Kepler\ mission demonstrated that small planets are abundant in the Milky Way Galaxy \citep{Borucki2010_Sci,Howard2012,Fressin2013,Petigura2013}.  What are the properties of small planets around nearby, bright stars, including their bulk and atmospheric compositions?  How do planet properties vary with stellar type and age?  The NASA \TESS\ mission is a two-year, all-sky survey that is finding small, transiting planets around nearby F,G,K, and M type stars \citep{Ricker2015}.  The all-sky strategy enables TESS to sample the transiting planets around brighter stars spanning a wider range of properties than were represented in the pencil-beam \Kepler\ survey.

A \TESS\ mission level-one science goal is to measure the masses of 50 sub-Neptune sized transiting planets\footnote{\href{https://heasarc.gsfc.nasa.gov/docs/tess/primary-science.html}{NASA TESS mission, accessed 2020 Aug 23}}.  The \TESS-Keck Survey (TKS) is a multi-institutional collaboration of Keck-HIRES users who are pooling Keck-HIRES time to meet this science goal and others \citep[see TKS-I,][and also TKS-0, Chontos et al. in prep.]{TKSI}.  The TKS science goals include determining the masses, bulk densities, orbits, and host star properties of planets in our survey.  Our survey targets were selected to answer broad questions about planet properties, formation, and evolution. 

TESS Object of Interest (TOI) 561 is a $V=10.2$ star that advances three of the TKS science goals: (1) to compare planetary siblings in systems with multiple transiting planets, (2) to characterize ultra-short period planets (USPs), and (3) to study planetary systems across a variety of stellar types.  Systems with multiple transiting planets provide excellent natural laboratories for testing the physics of planet formation, since the planets all formed around the same star and from the same protoplanetary disk.  TOI-561 is a bright star for which planet masses, interior compositions, and eventually atmospheric compositions can be determined through follow-up efforts.  Our investigation of TOI-561 advances our goal to compare the fundamental physical properties of small-planet siblings in extrasolar systems.

TOI-561 also hosts a USP that has an orbital period of  $<1 $ day and a radius consistent with a rocky composition \citep[e.g.,][]{Weiss2014,Rogers2015}\footnote{The definition of USPs as having $P<1$ day is somewhat arbitrary; see \citet{Sanchis-Ojeda2014} vs. \citet{Dai2018}.)}.  The present-day location of USPs corresponds to the former evacuated region of the protoplanetary disk.  Because the protoplanetary disk cavity forms during the first few million years of the star's existence, this inner region should have been depleted of the building blocks necessary to assemble planets.  Thus, the formation of USPs is poorly understood, but likely involves migration to overcome the low local density of solids.  Characterizing the mass and bulk density of TOI-561 b clarify how it and  other USPs formed.

We did not initially select TOI-561 for its host star properties, but we discovered during our investigation that TOI-561 is a member of the galactic thick disk.  Its low metallicity, high alpha abundance, and old age make it a special case that may advance our understanding of both multiplanet systems and the formation of USPs.  Its unusual chemistry, kinematics, and age also address a third goal of TKS, which is to study planetary systems across a variety of stellar types.

In \S\ref{sec:TESS}, we describe the TESS photometry, including the signals of the three transiting planet candidates.  In \S \ref{sec:stellar} we characterize the host star.  We describe our methods of planet candidate validation with ground-based photometry (\S\ref{sec:photometry}) and high-resolution imaging (\S\ref{sec:imaging}), and confirmation with radial velocities (\S\ref{sec:rvs}).  We describe the planet masses and densities in \S\ref{sec:masses}.  We discuss the planetary system orbital dynamics and prospects for future atmospheric characterization in \S\ref{sec:discussion}.  We conclude in \S\ref{sec:conclusion}.

\section{TESS Photometry \label{sec:TESS}}
\begin{figure*}
    \centering
    \includegraphics[width=\textwidth]{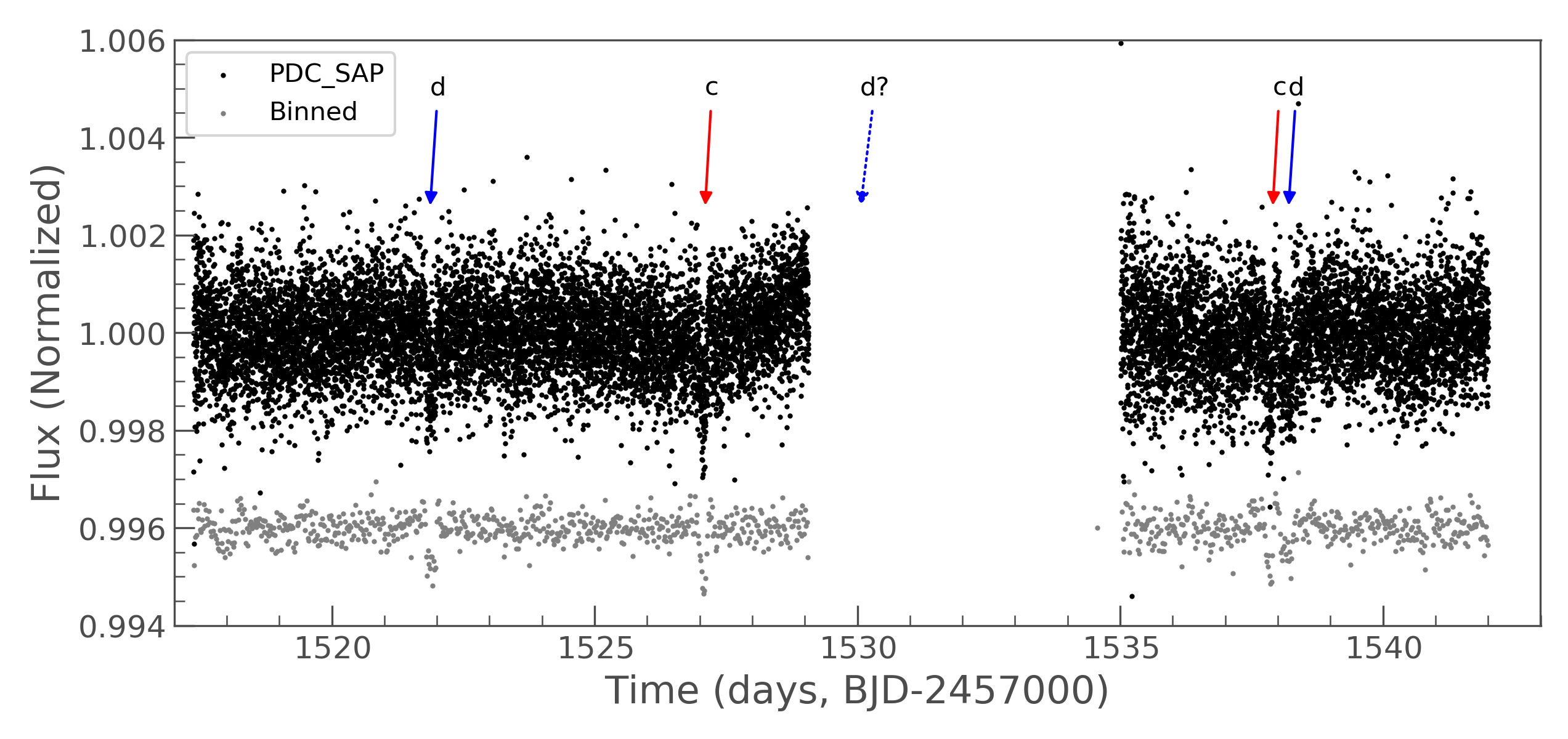}
    \caption{The presearch data conditioning SAP \textit{TESS} photometry of TOI-561 (black points) and the same photometry but binned every 13 data points and flattened with a Savitzky-Golay filter \citep[][gray points, with flux offset]{Savitzky1964}.  Individual transits of planets c (red arrows) and d (blue arrows) are marked.  A third transit of planet d could have occurred in the time series gap (blue dotted arrow).  A planet at $P=0.44$ days (planet b) is also present, but the transits are too shallow to see in these data (see Figures \ref{fig:transits} and \ref{fig:rvs}).}
    \label{fig:lightcurve}
\end{figure*}

\begin{figure}
    \centering
    \includegraphics[trim=190 10 30 0, clip, width=\columnwidth]{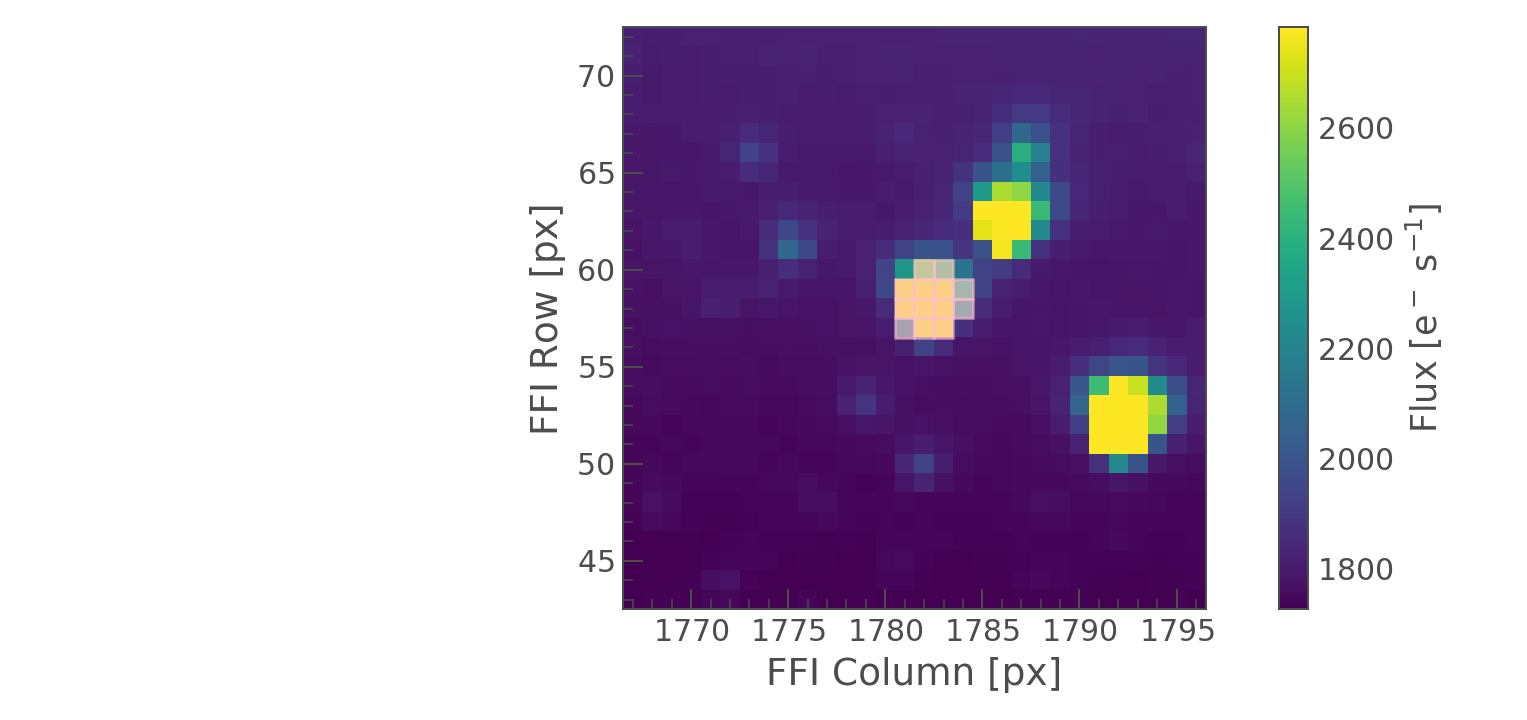}
    \caption{The \textit{TESS} Full-Frame Image centered on TOI-561.  The SPOC-defined aperture is a pale pink overlay on the central star.  The \textit{TESS} plate scale is 21\arcsec per pixel.  The target star has \textit{TESS} magnitude 9.49.  False positive scenarios in which the bright nearby stars are eclipsing binaries, with some flux contaminating the target pixels, are ruled out with the data validation centroid test (planet d) and/or follow-up ground based photometry (\S\ref{sec:photometry}, planets b and c).}
    \label{fig:aper}
\end{figure}

The vetting team of the TESS Science Processing Operations Center (SPOC) identified three transiting planets in their analysis of the photometry for TESS Input Catalog (TIC) ID 377064495 \citep{Jenkins2016,Twicken2018,Li2019}.  The presearch data conditioning simple aperture photometry (PDCSAP) is shown in Figure \ref{fig:lightcurve} \citep{Stumpe2014, Smith2012, Stumpe2012}.  The star was observed in Sector 8 at 2-minute cadence.  The SPOC-defined aperture is overlaid on the target in a Full Frame Image (FFI) in Figure \ref{fig:aper}.  The first planet candidate the SPOC pipeline detected is at $P=10.78$ days (TOI-561.01, planet c) based on two transits, with SNR 9.8.  After masking the flux near the transits of planet c, the SPOC pipeline detected a planet candidate at $P=0.45$ days (TOI-561.02, planet b) based on 55 transits, with SNR 10.0.  After masking the flux near transits of both planets c and b, the SPOC pipeline detected a planet candidate  at $P=16.4$ days (TOI-561.03, planet d), which transits twice, with SNR 9.2.

Several attributes of the TESS photometry complicate our analysis, particularly for planet d.  There is a gap partway through the time series that creates an alias in our interpretation of the transit signals.  The timing of the gap corresponds to a data download and also an unplanned interruption in communication between the instrument and spacecraft\footnote{\href{https://archive.stsci.edu/missions/tess/doc/tess_drn/tess_sector_08_drn10_v02.pdf}{TESS Data Release Notes: Sector 8, DR10}}.  The USP transited 55 times during the \textit{TESS} observations, leading to a robust ephemeris determination (although individual transits are too shallow to identify by eye in the photometry; see \S\ref{sec:photometry} for the phase-folded photometry and \S\ref{sec:rvs} for the RV planet confirmation).  However, only two transits of planet c and two transits of planet d were detected.  The transits of c and d occurred on different sides of the data gap.  For planet c, the non-detection of additional transits in the \textit{TESS} photometry leads to a robust determination of the orbital period at $10.78$ days, but for planet d, periods of 16 days (there is no transit during the gap) or 8 days (there is a transit in the gap, see the dotted blue arrow in Figure \ref{fig:lightcurve}).

Another challenge is that the second transit of planet d overlaps with a transit of planet b and is near a transit of planet c, and so much of the photometry during and near the second transit of d was masked in the original pipeline.  The lack of photometric continuum around the transit makes it difficult to isolate it and determine an accurate midtime, depth and duration.

To mitigate the frequent gaps from masking planets b and c, we ran a custom iteration of the SPOC DV pipeline.  We first identified planets c and b, but subtracted the best-fit models rather than masking the transits entirely so that we did not remove valuable continuum or in-transit data from the region with overlapping transits.  We then identified and fit planet d.  In our custom DV analysis, the depth for the odd transit of planet d is $947\pm126$ ppm and the depth for the even transit is $856\pm124$ ppm. The duration for the odd transit is $5.06\pm0.57$ hours and the duration for the even transit is $5.62\pm0.62$ hours.  The difference in the odd vs. even transit depths is 0.51$\sigma$ and the difference in the odd vs. even transit durations is 0.66$\sigma$. The transit depths from our custom DV analysis are consistent with the values from the original report.

A key difference between the original pipeline and our custom analysis is that the original SPOC pipeline identified the time between the two transits of planet d as 16.37 days, whereas in our custom analysis, that interval is 16.29 days.  The partial masking of planet d's transit in the original pipeline likely caused an inaccurate transit midpoint determination for the second transit, producing the inaccurate orbital period.  Our revised orbital period of 16.29 days implies that several follow-up photometric efforts for planet d were off by $>1$ day (\S\ref{sec:photometry}).

\begin{figure*}
    \centering
    \includegraphics[width=\textheight, angle=90]{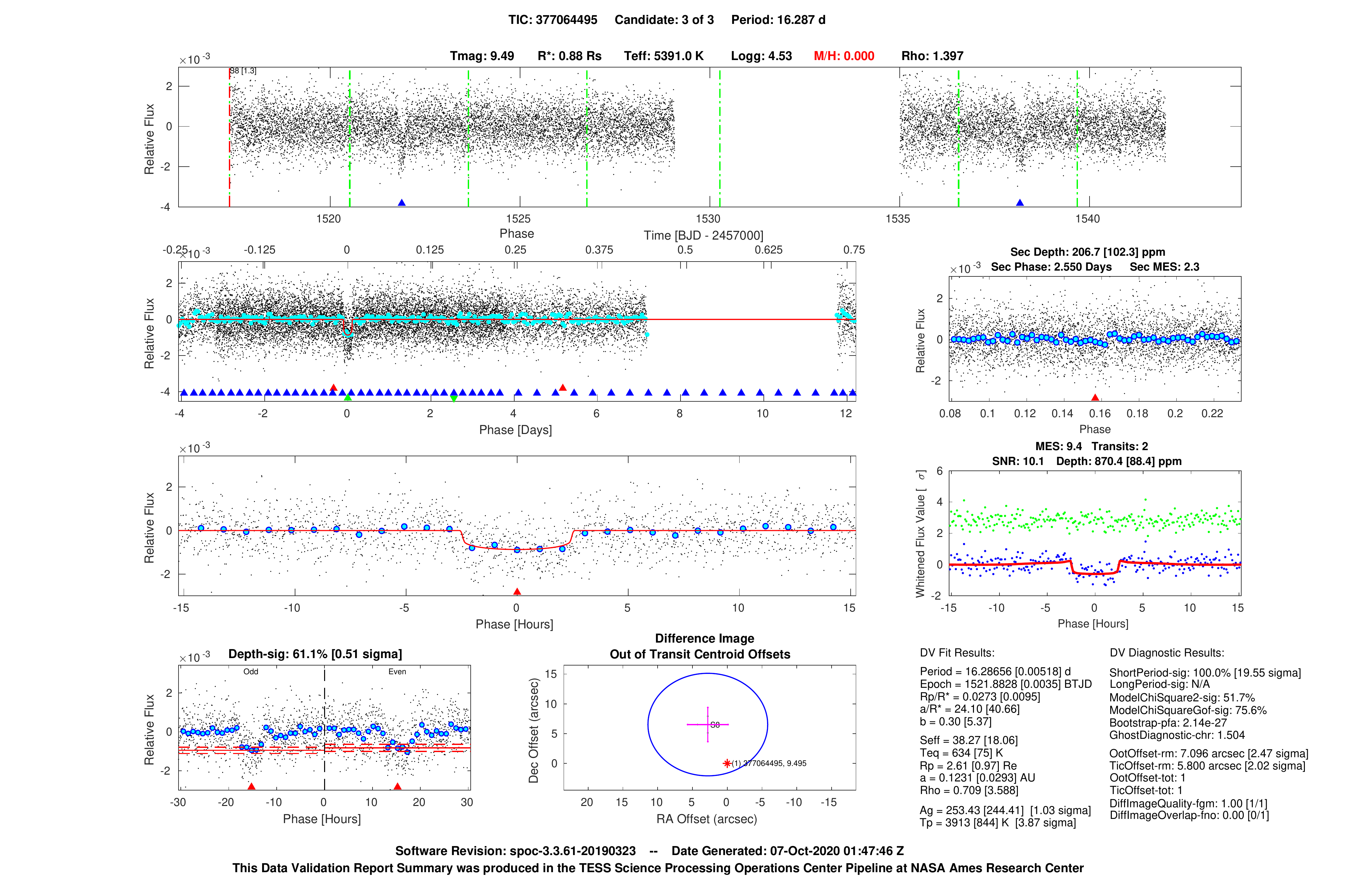}
    \caption{A custom run of the SPOC pipeline and DV analysis for TOI-561 provided more robust parameters than the default SPOC analysis. In the default analysis, the transits of planet d were affected by many short data gaps that resulted from masking the transits of b and c.  In our custom analysis, we subtracted (instead of masking) the transits of planets b and c, finding $P_d=16.29$ days (instead of 16.37 days).  The depths and durations of the two transits of planet d (bottom left panel) are consistent to $<1\sigma$.}
    \label{fig:custom_DV}
\end{figure*}

Despite the challenges related to planet d, the three planet candidates performed well in the data validation (DV) diagnostic tests.  The candidates passed all of the tests except for the difference image centroiding test, which placed the source for 561 b within 11\arcsec, 561 c within 23\arcsec, and 561 d within 7\arcsec (and a passing score for this test).  They all passed the ghost diagnostic test as well, indicating that if they were due to background eclipsing binaries, the offending star would have to be within a pixel of the location of the target star.  All three planet candidates pass the SPOC pipeline odd-even test, with insignificant differences between the depths of odd-numbered vs. even-numbered transits.\footnote{In the scenario of an 8 day period for planet d, the odd-even test is not meaningful, as the two transit-like events would correspond to two odd-numbered events.}

The combination of data validation and ground-based follow-up is sufficient to rule out a broad variety of astrophysical false positive scenarios for the planets.  Planet d passes the centroiding test in the DV report, ruling out an NEB as the source of the transits.  Through ground-based photometry, we recovered additional on-target transits of planet c, and also ruled out an NEB as the source of the transits for planet b (\S\ref{sec:photometry}).  Ground-based high-resolution imaging (\S\ref{sec:imaging} rules out background eclipsing binary false positives for all three planets.  Radial velocities (\S\ref{sec:rvs}) rule out that the target star itself is a spectroscopic eclipsing binary, and we detect planetary mass RV signals at the ephemerides of planets b and c.

\section{Stellar Properties \label{sec:stellar}}

\subsection{High-Resolution Spectroscopy}

We obtained a high signal-to-noise spectrum at $R=60,000$ (\S\ref{sec:rvs}) of \target\ to determine atmospheric parameters and detailed chemical abundances using the line list and forward modeling procedure of \citet{Brewer:2016_analysis}.  The modeling uses Spectroscopy Made Easy (SME) \citep{Valenti:1996_sme,Piskunov:2017_sme_evolution} in an iterative scheme that alternates between solving for global stellar properties and a detailed abundance pattern.  We begin by estimating \teff\ from B-V colors then fitting for \teff, \logg, [M/H], Doppler line broadening, and the abundances of the $\alpha$ elements calcium, silicon, and titanium.  All other elements are scaled solar values based on the overall metallicity given by [M/H] and the initial abundances are set to solar. The temperature of the resulting model is perturbed by $\pm 100\ \mathrm{K}$ and used as input to re-fit the spectrum.  The $\chi^2$ weighted average of the global stellar parameters are then fixed and used as the input for the next step of simultaneously fitting for the abundances of 15 elements.  

Simultaneous fitting of the elements is critical in obtaining precise abundances due to chemical processes in the stellar photosphere \citep[e.g., ]{Ting:2018_chemistry}.  The formation of molecules in cooler stars, even in very low numbers, can alter the atomic number densities and hence measured abundances using only isolated atomic lines.

The global parameters and abundance pattern obtained in the first iteration are then used as an initial guess for a second fitting following the same steps.  Finally, the macroturbulence is set using a \teff\ relation from \citet{Brewer:2016_analysis} and we solve for the projected rotational velocity, \vsini, with all other parameters fixed.  The resulting gravities have been shown to be consistent with those from asteroseismology to within 0.05 dex and the abundance uncertainties are between 0.01 - 0.04 dex \citep{Brewer2015}.  An empirical correction is applied to the abundances as a function of temperature \citep{Brewer:2016_analysis}, which adds additional uncertainty to the absolute abundance, especially at temperatures between 5000~K - 5500~K, and we adopt 0.05 dex uncertainty for most elements.  Our analysis yielded a low stellar metallicity and high alpha-abundance (\feh$=$\fehstar, [$\alpha/$Fe]$=+0.23\pm0.05$, see Table \ref{tab:stellar}).

The effective temperature derived from the SME analysis ($5326 \pm 25$\,K)\footnote{This error is the formal uncertainty, not the adopted error.} is in good agreement with alternative estimates using SpecMatch Synth \citep[$5249 \pm 110$\,K, ][]{Petigura2015PhD}, SpecMatch-Emp \citep[$5302 \pm 110$\,K,][]{Yee2017}, color-\teff\ relations applied in the \textit{TESS} Input Catalog \citep[$5440 \pm 110$ K,][]{stassun18} and applying a $J-K$ color-\teff\ relation \citep[$5300 \pm 110$ K,][]{casagrande10}. We adopted the SME-derived solution, with an error bar calculated from the standard deviation of \teff\ estimates from different methods: $5326\pm64$ K.

For each spectrum, we measure the Mt. Wilson S-value, an indicator of the chromospheric magnetic activity.  The Mt. Wilson S-value is a measure of the strength of the emission cores in the Ca II H and K lines relative to nearby continuum flux.  Our procedure for determining the S-values is described in \citet{Isaacson2010}.   See \S\ref{sec:rvs}, Table \ref{tab:rvs}, for the full S-value time series.  A Lomb-Scargle periodogram of the S-values results in peaks near 100 days and 230 days, neither of which is near the expected rotation period or magnetic activity cycle of this old K dwarf (\S\ref{sec:rvs}).

\subsection{Distance Modulus \& Isochrone Modeling}\label{sec:isochrone}

Stellar atmosphere and interior models are typically calculated using a solar-scaled $\alpha$ element abundance mixture and thus assume $\feh =  \mh$. To account for the non-solar $\alpha$ abundances of \target, we averaged the individual abundance measurements for $\rm{[Mg/H]}$, $\rm{[Si/H]}$, $\rm{[Ca/H]}$ and $\rm{[Ti/H]}$ to derive $\afeh = \afestar$, and then applied the calibration by \citet{salaris93} to convert the measured \feh\ value into an overall metal abundance, yielding $\rm{[M/H]} = -0.24 \pm 0.10$:
\begin{equation}
\mathrm{[M/H]} = \mathrm{[Fe/H]} + \mathrm{log_{10}}(0.694 \times 10^{\mathrm{[\alpha/Fe]}} + 0.306).
\end{equation}

We adopted a conservative uncertainty of 0.1\,dex for \mh\ to account for potential systematics in the \citet{salaris93} calibration.

Next, we used \teff, \mh, \logg, the \gaia\ DR2 parallax (adjusted for the $0.082 \pm 0.033$~mas zero-point offset for nearby stars reported by \citealt{stassun18b}), 2MASS K-band magnitude, a 3D dust map and bolometric corrections to calculate a luminosity by solving the standard distance modulus, as implemented in the ``direct mode'' of \texttt{isoclassify} \citep{huber17}. We then combined the derived luminosity with \teff\ and \mh\ to infer additional stellar parameters (mass, radius, density) using the ``grid mode'' of \texttt{isoclassify}, which performs probabilistic inference of stellar parameters using a grid of MIST isochrones \citep{choi16}. The isochrone-derived \logg\ ($4.50 \pm 0.03$\,dex) is in excellent agreement with spectroscopy ($4.52 \pm 0.05$\,dex), confirming that no additional iteration in the above steps is required for a self-consistent solution.  The derived age of the isochrone fit is $10\pm3$ Gyr, consistent with the mean age of a galactic thick disc star (see following section).

The full set of stellar parameters are listed in Table \ref{tab:stellar}. The results show that \target\ is an early-K dwarf with a radius of $\rstar=\radstar\,\rsun$ and mass $\mstar=\massstar\,\msun$. We note that the quoted uncertainties are formal error bars and do not include potential systematic errors due to the use of different model grids (Tayar et al. in prep.).  For example, the stellar radius in Table 1 is 3\% lower than predicted from an application of the Stefan-Boltzmann law using either the ``direct mode'' of \texttt{isoclassify} or SED fitting \citep{Stassun2017AJ}, both of which yield $0.86\pm0.02\,\rsun$. However, this 3\% ($\approx$\,1\,$\sigma$) difference does not significantly affect our main conclusions on the properties of the planets in the TOI-561 system, since the planet density errors are dominated by uncertainties in the planet masses (see \S\ref{sec:masses}).

We used the stellar evolution model fitting tool \texttt{kiauhoku} \citep{Claytor2020} to estimate the rotation period of TOI-561.  Using the stellar \teff, \feh, and [$\alpha$/Fe] from Table \ref{tab:stellar} as inputs, and assuming an age of $10\pm3$ Gyr, we found two different model-dependent estimates of the rotation period.  Assuming the magnetic braking law described in \citet{vanSaders2013}, we found $P_\mathrm{rot} = 38.5\pm7.3$ days, but assuming the stalled-braking law of \citet{vanSaders2016}, we found $P_\mathrm{rot} =35.7\pm3.4$ days.  These rotation periods are consistent with the upper limit of \vsini\ we determined spectroscopically.  However, the estimated rotation periods differ significantly from the periodicity identified in the Mt.\ Wilson S-value activity indices (see Table \ref{tab:rvs}), suggesting that the rotation period is not detected in the S-value time series.  A rotation period of $>30$ days is likely too long to identify in the single sector of TESS photometry.  We checked the archives of several ground-based photometric surveys, but TOI-561 saturates in ASAS-SN and Pan-STARRS, and it is too close to the equator to be included in WASP.

\begin{table}
\begin{center}
\caption{Host Star Characteristics\label{tab:stellar}
}
\renewcommand{\tabcolsep}{0mm}
\begin{tabular}{l c}
\tableline\tableline
\multicolumn{2}{c}{Basic Properties} \\
\noalign{\smallskip}
\hline
Tycho ID & 243-1528-1 \\
TIC ID & 377064495 \\
Gaia DR2 ID & 3850421005290172416 \\
Right Ascension	& 09:52:44.44 \\
Declination	& +06:12:57.00 \\
Tycho $V_T$ Magnitude & 10.25 \\
\tess\ Magnitude & 9.49  \\
2MASS $K$ Magnitude & 8.39 \\
\hline
\multicolumn{2}{c}{\gaia\ DR2 Astrometry} \\
\hline
Parallax, $\pi$ (mas) & $11.627 \pm 0.067$  \\
Radial Velocity (km/s) & $79.54 \pm 0.56$  \\
Proper Motion in RA (mas/yr) & $-108.432 \pm 0.088$  \\
Proper Motion in DEC (mas/yr) & $-61.511 \pm 0.094$  \\
\hline
\multicolumn{2}{c}{High-Resolution Spectroscopy} \\
\hline
Effective Temperature, \teff\, (K) & \teffstar \\
Surface Gravity, \logg\ (cm s$^{-2}$) & 4.52 $\pm$ 0.05\\
Projected rotation speed, \vsini\ (km\,s$^{-1}$) & < 2.0 \\
log$R^{\prime}_\mathrm{HK}$ (dex) & -5.1 \\
Iron Abundance, [Fe/H] (dex) & \fehstar \\
Carbon Abundance, [C/H] (dex) & $-0.19\pm0.05$\\
Nitrogen Abundance, [N/H] (dex) & $-0.51\pm0.05$\\
Oxygen Abundance, [O/H] (dex) & $+0.09\pm0.05$\\
Sodium Abundance, [Na/H] (dex) & $-0.39\pm0.05$\\
Magnesium Abundance, [Mg/H] (dex) & $-0.20 \pm 0.05$ \\
Aluminum Abundance, [Al/H] (dex) & $-0.19 \pm 0.05$ \\
Silicon Abundance, [Si/H] (dex) &  $-0.24 \pm 0.05$ \\
Calcium Abundance, [Ca/H] (dex) &  $-0.27 \pm 0.05$ \\
Titanium Abundance, [Ti/H] (dex) &  $-0.20 \pm 0.05$ \\
Vanadium Abundance, [V/H] (dex) &  $-0.27 \pm 0.05$ \\
Chromium Abundance, [Cr/H] (dex) &  $-0.43 \pm 0.05$ \\
Manganese Abundance, [Mn/H] (dex) &  $-0.60 \pm 0.05$ \\
Nickel Abundance, [Ni/H] (dex) &  $-0.37 \pm 0.05$ \\
Yttrium Abundance, [Y/H] (dex) &  $-0.42 \pm 0.05$ \\
Alpha Abundance, [$\alpha$/Fe] (dex) & \afestar \\
\hline
\multicolumn{2}{c}{Distance Modulus \& Isochrone Modeling} \\
\hline
Stellar Luminosity, \lstar\ ($\lsun$) & $0.522 \pm 0.017$ \\
Stellar Mass, \mstar\ (\msun)& \massstar \\
Stellar Radius, \rstar\ (\rsun)& \radstar \\
Stellar Density, \rhostar\ ($\rho_\sun$)& \denstar \\
Surface Gravity, \logg\ (cgs) & \loggseis \\
Age (Gyr) & $10\pm3$ \\
\hline
\multicolumn{2}{c}{Transit Modeling} \\
\hline
Limb Darkening ({\it TESS} band), $q_1$ & $0.2^{+0.2}_{-0.2}$ \\
Limb Darkening ({\it TESS} band), $q_2$ & $0.4^{+0.3}_{-0.2}$ \\
\end{tabular}
\end{center}
\flushleft Notes: The \tess\ magnitude is adopted from the \textit{TESS} Input Catalog \citep{stassun18}, and the kinematics are taken from Gaia DR2 \citep{lindegren18}.  Stellar parameters from isochrone modeling are formal uncertainties only, and do not incorporate systematic errors from different model grids.  The transit modeling is described in Section \ref{sec:photometry}.
\end{table}

\subsection{Galactic Evolution}

Early studies of star counts in the Milky Way revealed two distinct populations in the galactic disk which dominate at different scale heights, commonly denoted the ``thin'' and ``thick'' disk population \citep{gilmore83}. Spectroscopic and photometric surveys have  shown that these populations can be approximately separated based on kinematics and chemical abundances, with thick disk stars being kinematically hotter \citep[e.g.][]{fuhrmann98}, older \citep{bensby05}, more metal-poor and enriched in $\alpha$ process elements \citep[e.g.][]{fuhrmann98}. 

The formation of the thick disk is still debated, with scenarios including external processes such as the accretion of stars from the disruption of a satellite galaxy \citep[e.g.][]{abadi2003} and induced star formation from mergers with with other galaxies \citep[e.g.][]{brook04}, or a natural dynamical evolution of our galaxy including radial migration \citep{schonrich09a,schonrich09b}. While the mere existence of a distinct thick disk is still in question \citep{bovy12},  spectroscopic and asteroseismic surveys have confirmed that chemically-identified ``thick disk'' stars belong to the old population of our galaxy, with typical ages of $\sim$\,11\,Gyr \citep{silva18}. 

The detection of exoplanets around different galactic stellar populations can provide powerful insights into their formation and evolution \citep{adibekyan12}. For example, the discovery of five sub-Earth sized planets orbiting the thick-disk star Kepler-444 \citep{campante15} demonstrated for the first time that terrestrial planet formation has occurred for at least $\sim$11\,Gyr, and the discovery of a close M dwarf binary companion demonstrated that this process can even proceed in a truncated protoplanetary disk \citep{dupuy16}. While \textit{TESS} probes nearby stellar populations it has significant potential to expand this sample. Indeed, \citet{gan20} recently presented the first \textit{TESS} exoplanet orbiting a thick disk star identified based on kinematics.

\begin{figure}
\begin{center}
\resizebox{\hsize}{!}{\includegraphics{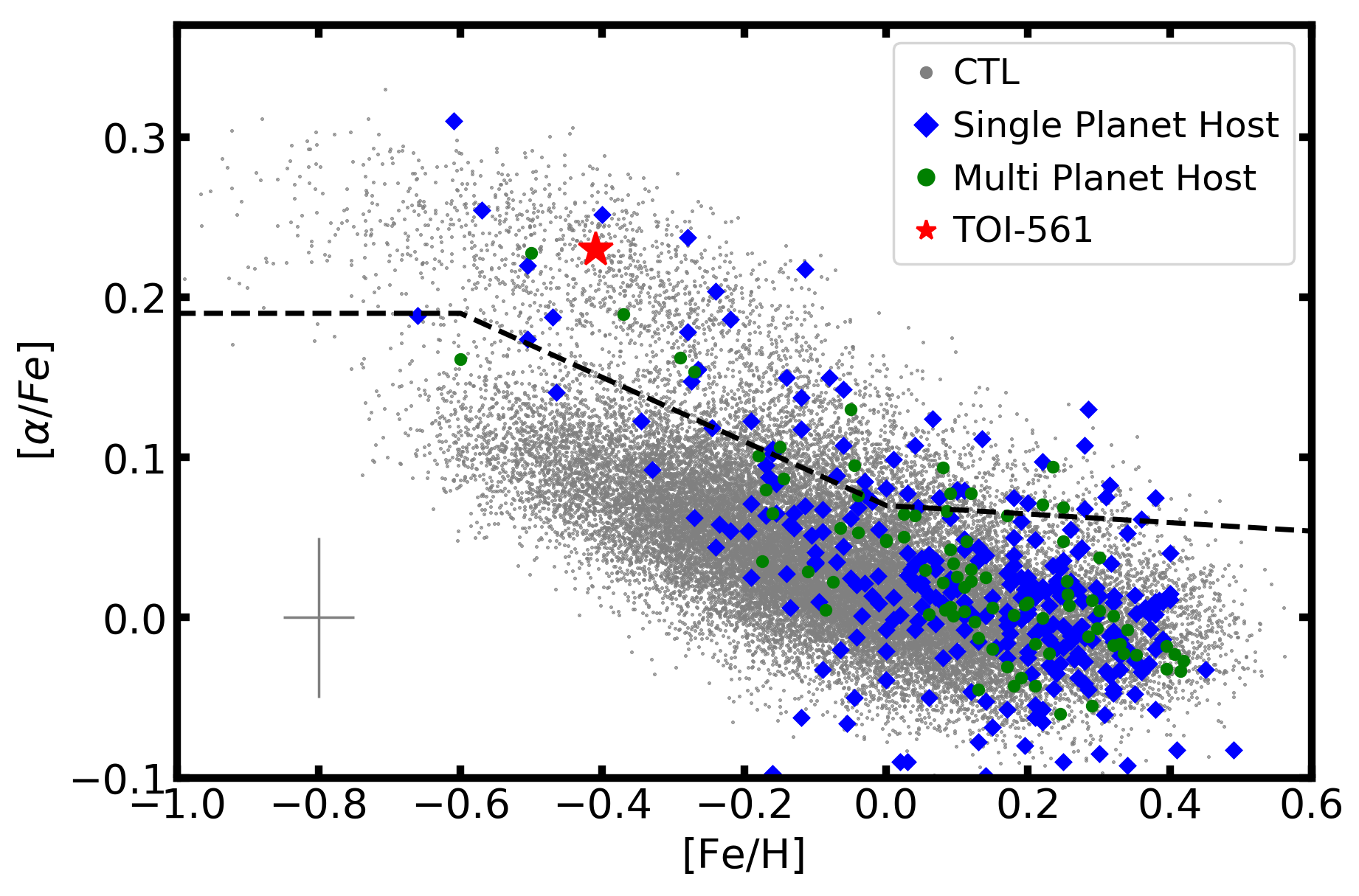}}
\caption{Iron abundance versus [$\alpha$/Fe] for stars in the \textit{TESS} candidate target list (CTL) observed by the GALAH survey \citep{desilva15,sharma18} and a sample of exoplanet host stars taken from Hypatia catalog \citep{hinkel14}. Known exoplanet hosts are separated into those with a single known planet (blue diamonds) and multiple known planets (green circles). The position of \target\ is marked by a red star. The black dashed line approximately separates the galactic thin disk and thick disk populations}
\label{fig:chemo}
\end{center}
\end{figure}

Figure \ref{fig:chemo} compares the chemical properties of \target\ with a sample of field stars in the \textit{TESS} candidate target list (CTL) observed by the GALAH survey \citep{desilva15,sharma18} and a sample of known exoplanet hosts from the Hypatia catalog \citep{hinkel14}. We calculated \afeh\ for stars in the Hypatia catalog in the same manner as for \target\, and discarded stars with abundance uncertainties $>0.2$\,dex (calculated from the scatter between different methods). \target\ is consistent with the thick disk in terms of its chemical abundances, in agreement with the high proper motions measured by Gaia (Table \ref{tab:stellar}) and the kinematic classification of \target\ by \citet{carillo20}. 
To independently confirm the kinematic classification, we used the  $UVW$ velocity vector of \target\ via the online velocity calculator of \citet{rodriguez:2016}, finding  $(U,V,W) = (-60.0, -70.9, +16.7)$~km~s$^{-1}$.  Using the probabilistic framework of \citet{bensby:2004} and \citet{bensby:2014} we find a thick-to-thin disk probability ratio of $TD/D=19$, indicating strong evidence that this star is a member of the thick disk.

\target\ is the first chemically and kinematically confirmed thick-disk exoplanetary system discovered by \textit{TESS}, the fifth known thick disk star known to host multiple planets, and the first thick-disk star known to host an ultra-period short planet. This further demonstrates that (1) small, rocky planets can form in metal-poor environments \citep[consistent with ][]{buchhave12}, (2) USPs are not tidally destroyed around old stars \citep[consistent with ][]{Hamer2020}, and (3) rocky planets have been forming for nearly the age of the universe.

\section{Time-Series Photometric Follow-up and Analysis \label{sec:photometry}}

We acquired ground-based time-series follow-up photometry of TOI-561 as part of the \textit{TESS} Follow-up Observing Program (TFOP)\footnote{\href{https://tess.mit.edu/followup}{TFOP website}} to attempt to (1) rule out nearby eclipsing binaries (NEBs) as potential sources of the \textit{TESS} detections and (2) detect the transits on target to refine the \textit{TESS} ephemerides. We used the {\tt TESS Transit Finder}, which is a customized version of the {\tt Tapir} software package \citep{Jensen:2013}, to schedule our transit observations. 

\subsection{LCOGT}
We observed TOI-561 using the Las Cumbres Observatory Global Telescope (LCOGT) 1-m networks \citep{Brown:2013} in Pan-STARRS $z$-short (zs) band. The telescopes are equipped with $4096\times4096$ SINISTRO cameras having an image scale of 0$\farcs$389 pixel$^{-1}$ resulting in a $26\arcmin\times26\arcmin$ field of view. The images were calibrated using the standard LCOGT BANZAI pipeline \citep{McCully:2018}, and the photometric data were extracted using the {\tt AstroImageJ} ({\tt AIJ}) software package \citep{Collins:2017}. A full transit window of TOI-561 b was observed continuously for 205 minutes on 19 April 2019 UT from the LCOGT Siding Spring Observatory (SSO) node. TOI-561 c was observed continuously for 381 minutes on 03 February 2020 UT from the LCOGT McDonald Observatory node and again on 17 March 2020 UT from the LCOGT Cerro Tololo Inter-American Observatory (CTIO) node for 230 minutes and then later on the same epoch from the LCOGT SSO node for 269 minutes. TOI-561 d was observed continuously for 300 minutes on 24 April 2020 UT from the LCOGT SSO node.

\subsection{NGTS} \label{sec:ngts}

The Next Generation Transit Survey \citep[NGTS;][]{wheatley2018ngts}, located at ESO's Paranal Observatory, is a photometric facility dedicated to hunting exoplanets. NGTS consists of twelve independently operated 20\,cm diameter robotic telescopes, each with an 8 square-degree field-of-view and a plate scale of 5\,\arcsec\,pixel$^{-1}$. The NGTS telescopes also benefit from sub-pixel guiding afforded by the DONUTS auto-guiding algorithm \citep{mccormac13donuts}. By using multiple NGTS telescopes to simultaneously observe the same star, NGTS can achieve ultra-high precision light curves of exoplanet transits \citep{bryant2020multicam, smith2020multicam}.

TOI-561 was observed on two nights using NGTS multi-telescope observations. On UT 2020 February 02 a full transit of TOI-561c was observed using three NGTS telescopes. A predicted transit ingress of TOI-561d was observed on the night UT 2020 March 05 using four NGTS telescopes. A total of 5179 images were obtained during the first observation, and 7791 during the second. Both sets of observations were performed using an exposure time of 10\,s and the custom NGTS filter (520 - 890\,nm). The airmass of the target was kept below 2 and the sky conditions were good for all the observations.

We reduced the NGTS images using the custom aperture photometry pipeline detailed in \citet{bryant2020multicam}. This pipeline performs source extraction and photometry using the \texttt{SEP} library \citep{bertin96sextractor, Barbary2016}. The pipeline also uses GAIA DR2 \citep{GAIA, GAIA_DR2} to automatically identify a selection of comparison stars which are similar to TOI-561 in terms of brightness, colour and CCD position.

\subsection{MuSCAT2}
We observed full transit windows of TOI-561 b continuously for 120 minutes on 23 April 2019 UT and 24 May 2020 UT simultaneously in $g$, $r$, $i$, and $z_\mathrm{s}$ bands with the MuSCAT2 multi-color imager \citep{Narita2018} installed at the 1.52~m Telescopio Carlos Sanchez (TCS) in the Teide Observatory, Spain. The photometry was carried out using standard aperture photometry calibration and reduction steps with a dedicated MuSCAT2 photometry pipeline, as described in \citet{Parviainen2020}. 

\subsection{PEST}
We observed a full transit window of TOI-561 b continuously for 205 minutes on 22 April 2019 UT in $\rm R_c$ band from the Perth Exoplanet Survey Telescope (PEST) near Perth, Australia. The 0.3 m telescope is equipped with a $1530\times1020$ SBIG ST-8XME camera with an image scale of 1$\farcs$2 pixel$^{-1}$ resulting in a $31\arcmin\times21\arcmin$ field of view. A custom pipeline based on {\tt C-Munipack}\footnote{http://c-munipack.sourceforge.net} was used to calibrate the images and extract the differential photometry.

\subsection{El Sauce}
We observed a full transit window of TOI-561 b continuously for 206 minutes on 23 April 2019 UT in $\rm R_c$ band from El Sauce Observatory in Coquimbo Province, Chile. The 0.36 m Evans telescope is equipped with a $1536\times1024$ SBIG STT-1603-3 camera with an image scale of 1$\farcs$47 pixel$^{-1}$ resulting in a $18.8\arcmin\times12.5\arcmin$ field of view. The photometric data were extracted using {\tt AIJ}.

\subsection{TOI-561 b}
The TOI-561 b SPOC pipeline transit depth is generally too shallow (290 ppm) for ground-based detection, so we checked all three stars within $2.5\arcmin$ that are bright enough to have caused the SPOC detection (i.e. \textit{TESS} magnitude < 18.1) for a possible NEB that could be contaminating the SPOC photometric aperture.  We estimate the expected NEB depth in each neighboring star by taking into account both the difference in magnitude relative to TOI-561 and the distance TOI-561 (to estimate the fraction of the star's flux that would be contaminating the TESS aperture for TOI-561). If the RMS of the 5-minute binned light curve of a neighboring star is more than a factor of 3 smaller than the expected NEB depth, we consider an NEB to be ruled out. We also visually inspect each star's light curve to ensure that there is no obvious eclipse-like signal, even though the RMS to estimated NEB depth threshold is met. Using a combination of the LCOGT, MuSCAT2, PEST, and El Sauce TOI-561 b follow-up observations, we rule out the possibility of a contaminating NEB at the SPOC pipeline ephemeris.

\subsection{TOI-561 c}
\begin{figure}
    \centering
    \includegraphics[trim=0 0 35 0, clip, width = 0.9\columnwidth]{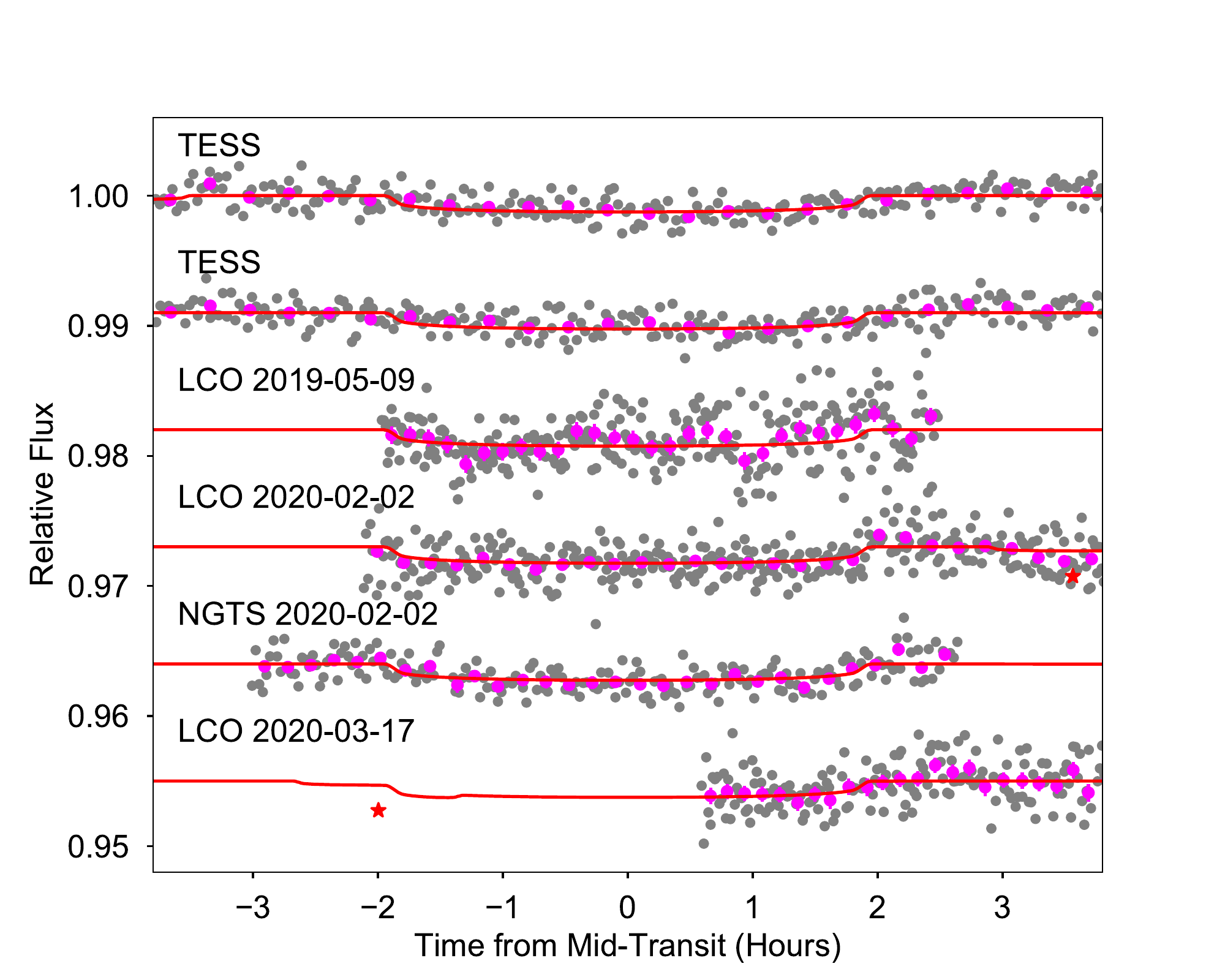}
    \caption{The individual transits of planet c from {\it TESS} and ground-based facilities. The magenta points are binned data. The red stars indicate the mid-transits of planet b which due to the short orbital period often overlap with the transits of planet c and d. The red solid is the best fit global model that include both {\it TESS} and ground-based photometry and model transits of all three planets simultaneously.  Note that the ground-based transits of planet d were acquired at times that are not consistent with our best-fit ephemeris, and they are not shown.}
    \label{fig:transits_individual}
\end{figure}

In the LCOGT observation of TOI-561 c on 03 February 2020 UT, we detected a 142 min early (0.3$\sigma$) $\sim1100$~ppm egress, relative to the nominal SPOC ephemeris, in a $9\farcs7$ radius aperture around the target star, which is not contaminated with any known Gaia DR2 stars. NGTS observed and detected the same transit on time (Fig. \ref{fig:transits_individual}). As a result, we revised the follow-up orbital period to 10.778325 days for further scheduling. The 17 March 2020 UT LCOGT CTIO and SSO observations then detected an on-time ingress and egress, respectively, at the revised ephemeris.

\subsection{TOI-561 d}
The LCOGT observation of TOI-561 d on 24 April 2020 UT covered an egress $\pm150$~minutes relative to the nominal SPOC pipeline ephemeris and provides $\sim1\sigma$ coverage of the original SPOC ephemeris uncertainty on the epoch of observation, but is 2 days away from the transit midpoint predicted by our revised 16.29 day ephemeris.  The LCOGT light curve does not show an obvious 923 ppm egress during the limited observation window. However, ingress- or egress-only coverage of transits with depths less than 1000 ppm from the LCOGT 1-m network can be difficult to interpret and reliably model due to potential trends in the data that may inject or mask a shallow ingress or egress. The NGTS observation on 05 March 2020 shows what might be the ingress of a transit of planet d. However, given the low SNR, this is not a high confidence detection.  Furthermore, if our revised 16.29 day ephemeris is correct, this observation was 2 days away from the transit midpoint.

\subsection{Transit Modeling}
\begin{figure}
    \centering
    \includegraphics[trim=15 45 35 110, clip, width = 0.9\columnwidth]{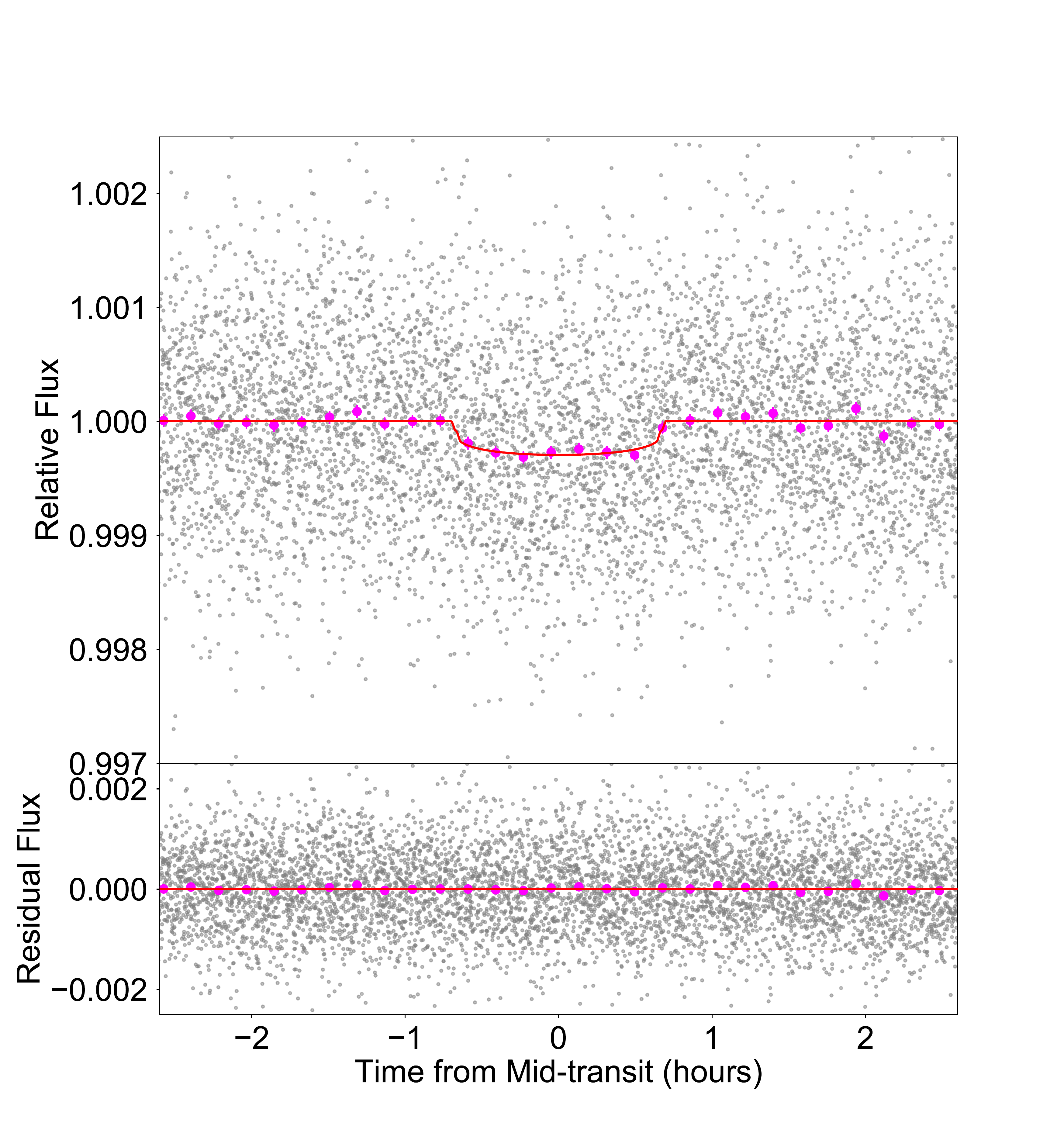}
    \includegraphics[trim=15 45 35 110, clip, width = 0.9\columnwidth]{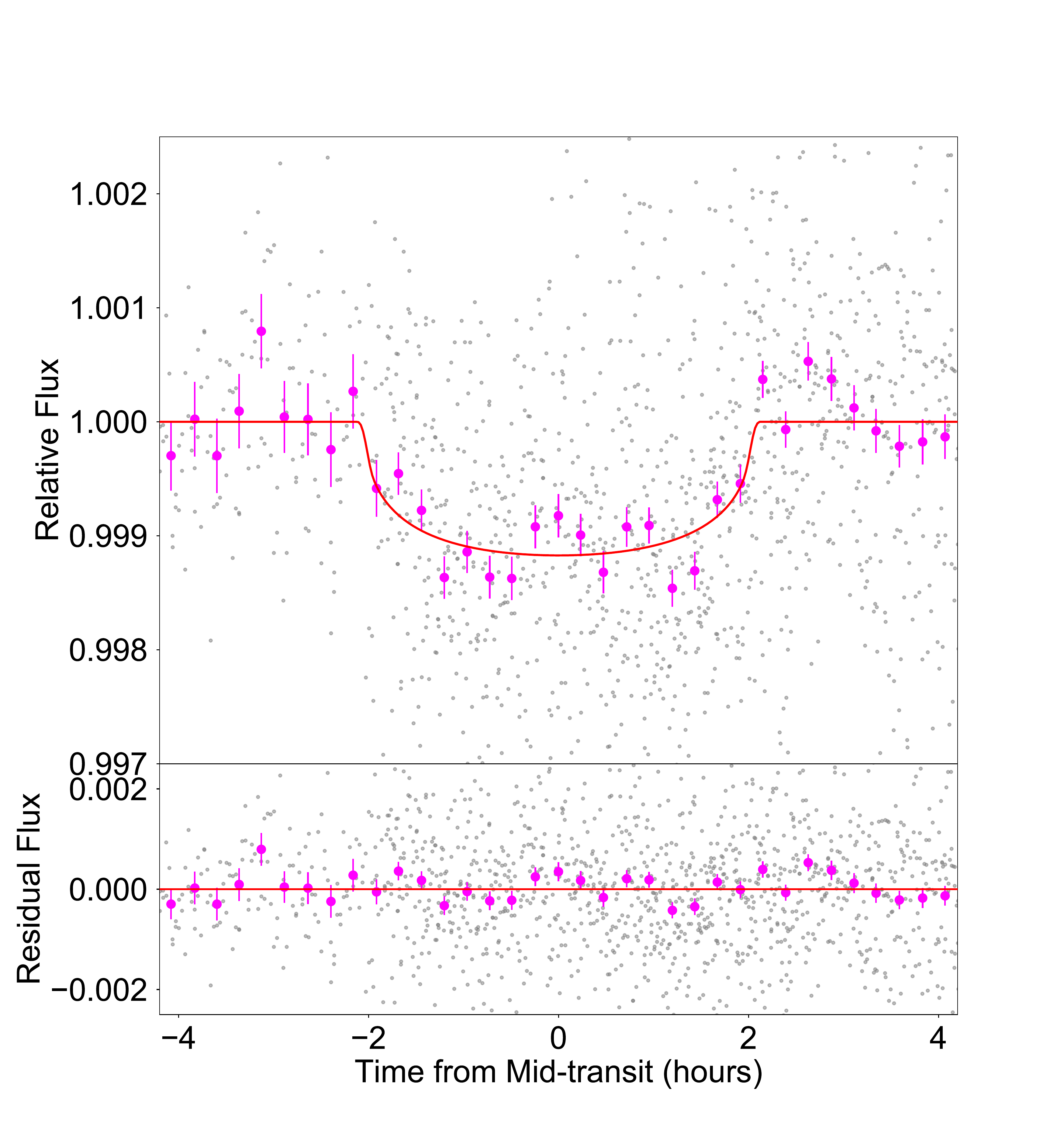}
    \includegraphics[trim=15 45 35 110, clip, width = 0.9\columnwidth]{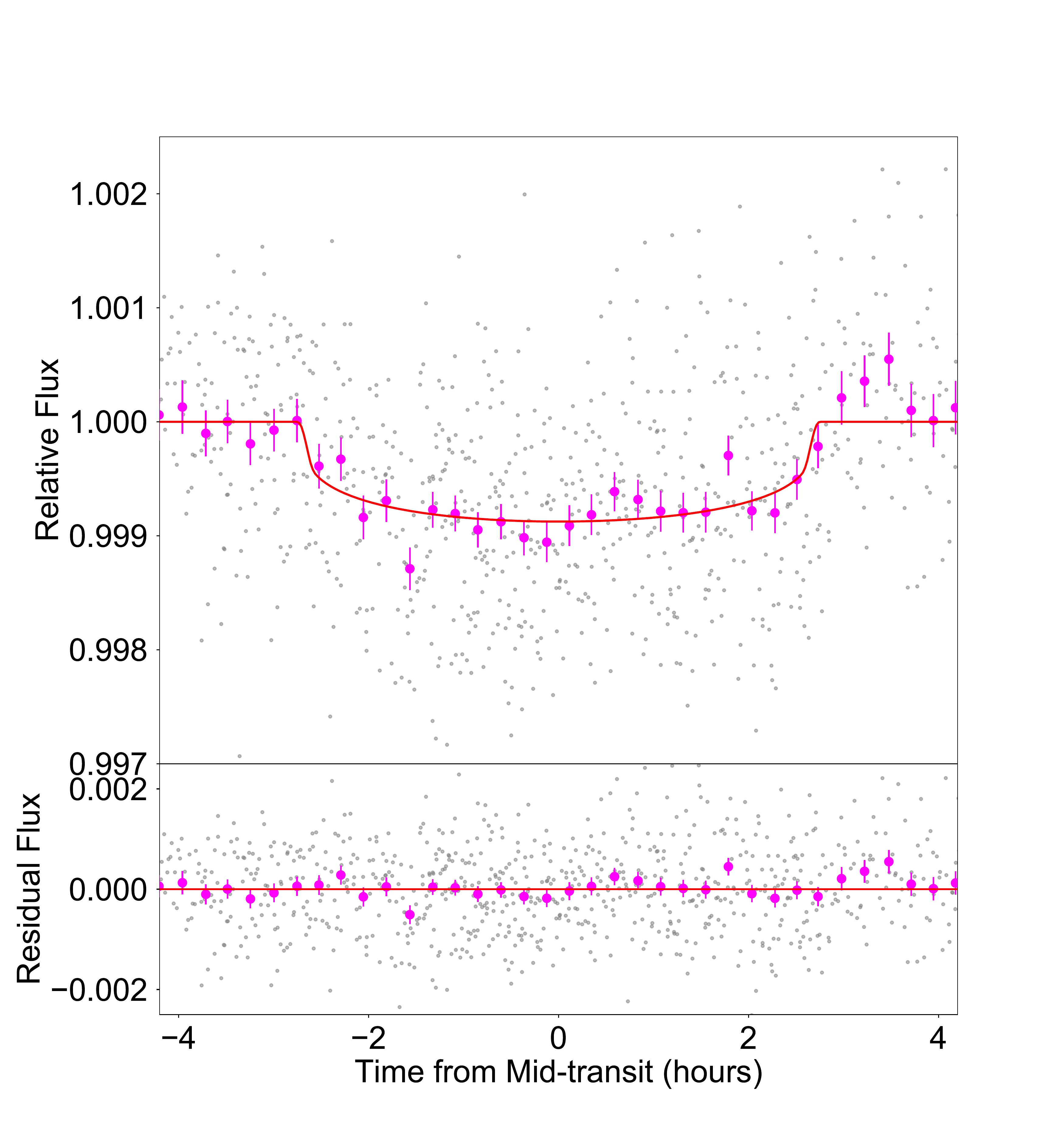}
    \caption{The phase-folded transits of TOI-561 planets b (top), c (middle), and d (bottom). We removed regions where two planets transit simultaneously before plotting. The magenta points show binned fluxes. The red solid line shows the best-fit transit models.}
    \label{fig:transits}
\end{figure}

Here we perform a joint analysis of \textit{TESS} light curve and ground-based follow-up to refine the planetary parameters. We downloaded the {\it TESS} light curve from the Mikulski Archive for Space Telescopes (MAST, Figure \ref{fig:lightcurve}). We isolated the transits of each planet with a window of three times the transit duration. We removed long term stellar variability/instrumental effect by fitting a cubic spline with 1.5-day knot length to the light curve after removing the transits. We also downloaded the ground-based follow-up observations from the \href{https://exofop.ipac.caltech.edu/tess/}{ExoFOP website}.   We used the \texttt{BATMAN} \citep{Kreidberg_Batman} package for transit modeling, using the transit ephemerides reported by the \textit{TESS} team as an initial guess for our model.   Our model uses the mean stellar density $\rho_\star$ as a global parameter (with a Gaussian prior as derived in Section \ref{sec:stellar} and Table \ref{tab:stellar}) on all three planets. For each planet, we allowed the radius ratio $R_p/R_\star$, the impact parameter $b$, the orbital period $P$, and the mid-transit time $T_c$ to vary freely. We assumed circular orbits for all three planets.  The mean stellar density $\rho_\star$ and the orbital period $P$ together constrain the scaled semi-major axes $a/R_\star$ of each planet. We adopted a quadratic limb-darkening law as parameterized by \citet{Kipping2013}. We allowed the coefficients $q_1$, $q_2$ to vary in different photometric bands. We then performed a Monte Carlo Markov Chain analyses with the \texttt{Python} package \texttt{emcee} \citep{Foreman-Mackey2013} to sample the posterior distribution of the various transit parameters. The results are summarized in Table \ref{tab:planet}, and Figure \ref{fig:transits} shows the best-fit transit models.

We found that the transit duration for d is long compared to what is expected for a planet in a circular, 16-day orbit, given the well-characterized stellar density.\footnote{The custom DV report, which does not use spectroscopic stellar parameters as priors, finds $\rhostar=0.7\,\rhosun$, which differs substantially from our spectroscopic determination of $\rhostar=\denstar\,\rhosun$.}  The long transit duration can be resolved with a moderate eccentricity for planet d.  We applied the photo-eccentric method of \citet{Dawson2012} to estimate the eccentricity of planet d, finding $e_d = 0.24^{+0.27}_{-0.13}$.  However, this eccentricity would likely result in instability of the 10 and 16 day planets, in contrast to the stable orbits we find assuming low eccentricities (see \S\ref{sec:discussion}).  The too-long duration of planet d is only worsened if we assume the $P_d=8$ day solution.  Another resolution of the long-duration transits of planet d is suggested in a contemporaneous paper by \citet{Lacedelli2020}: the two apparent transits of planet d might be single-transit events from two distinct planets, each with $P > 16$ days.  The similar depths of the transits could result from the observed ``peas in a pod'' pattern, wherein planets in the same system often have similar sizes \citep{Weiss2018}.  We do not detect a secure RV signal at $P>16$ days that would be consistent with the orbit of either such planet (\S\ref{sec:rvs}).  Yet another possibility is that the apparent 2$\sigma$ tension between the transit duration and orbital period is the result of systematic or random errors in the photometry.  Ultimately, additional high-precision photometry is needed to test these various possible explanations for the long durations of planet d.  For the rest of this paper, we will assume that the transits are due to a 16 day planet (except where stated otherwise).

\section{High-Resolution Imaging \label{sec:imaging}}
As part of our standard process for validating transiting exoplanets to assess the the possible contamination of bound or unbound companions on the derived planetary radii \citep{ciardi2015} and search for possible sources of astrophysical false positives (e.g., background eclipsing binaries), we obtained high-angular resolution imaging in the near-infrared and optical. 

\subsection{Gemini-North and Palomar}
We utilized both Gemini-North with NIRI \citep{Hodapp2003} and Palomar Observatory with PHARO \citep{hayward2001} to obtain near-infrared adaptive optics imaging of TOI~561, on 24-May-2019 and 08-Jan-2020 respectively. Observations were made in the Br$\gamma$ filter $(\lambda_o = 2.1686; \Delta\lambda = 0.0326\mu$m).  For the Gemini data, 9 dithered images with an exposure time of 2.5s each were obtained; at Palomar, 15 dithered frames with an exposure of 2.8s each were obtained. In both cases, the telescope was dithered by a few arcseconds between each exposure, and the dithered science frames were used to create a sky background. Data were reduced using a custom pipeline: we removed bad pixels, performed a sky background subtraction and a flat correction, aligned the stellar position between images and coadded. The final resolution of the combined dithers was determined from the full-width half-maximum of the point spread function; 0.13\arcsec\ and 0.10\arcsec\ for the Gemini and Palomar data, respectively.

The sensitivities of the final combined AO images were determined by injecting simulated sources azimuthally around the primary target every $20^\circ $ at separations of integer multiples of the central source's FWHM \citep{furlan2017, lund2020}. The brightness of each injected source was scaled until standard aperture photometry detected it with $5\sigma $ significance. The resulting brightness of the injected sources relative to the target set the contrast limits at that injection location. The final $5\sigma $ limit at each separation was determined from the average of all of the determined limits at that separation and the uncertainty on the limit was set by the rms dispersion of the azimuthal slices at a given radial distance. 

\begin{figure*}
    \centering
    \includegraphics[width=0.4\textwidth]{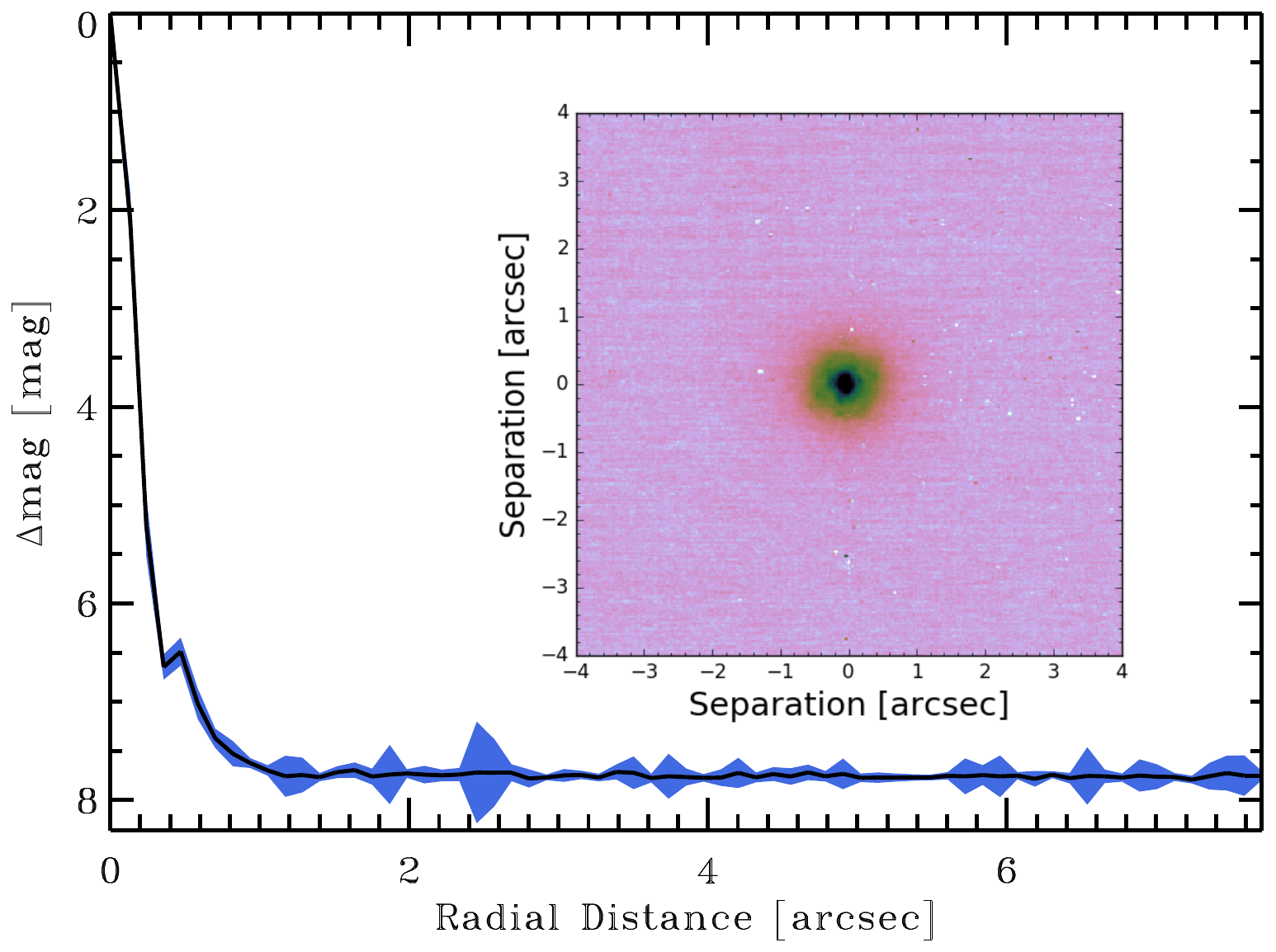}
    \includegraphics[width=0.45\textwidth]{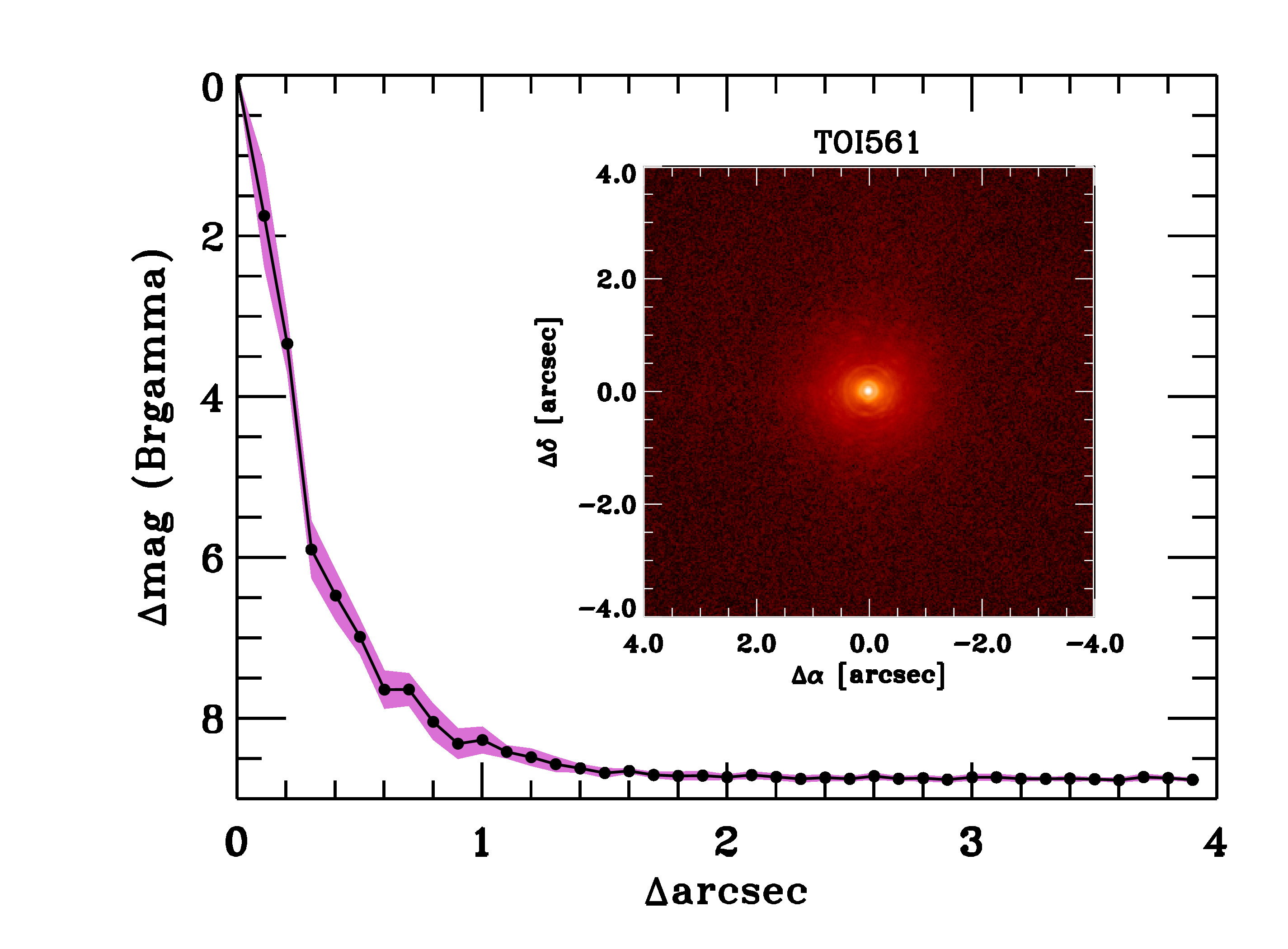}
    \includegraphics[width=0.4\textwidth]{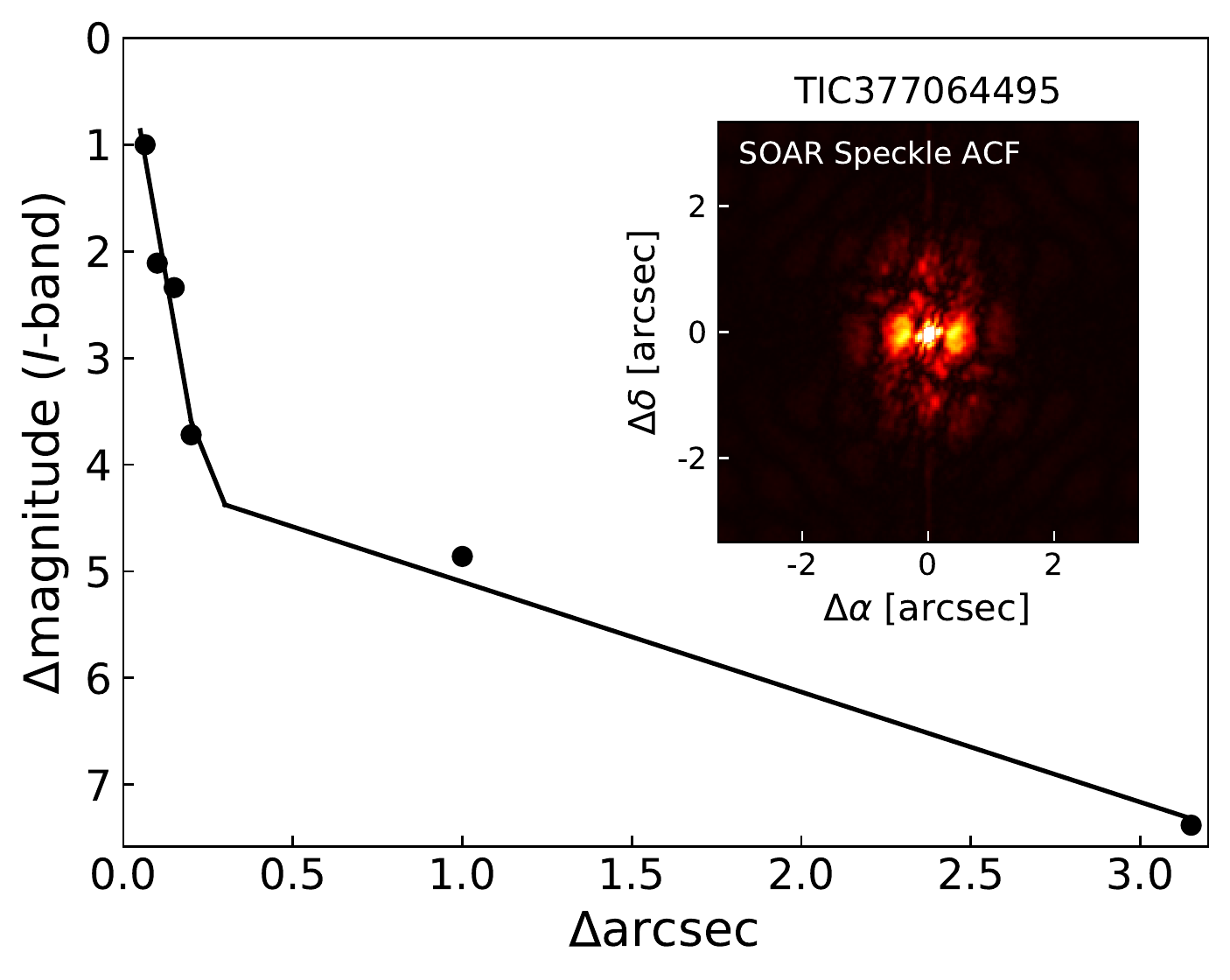}
    \includegraphics[width=0.4\textwidth]{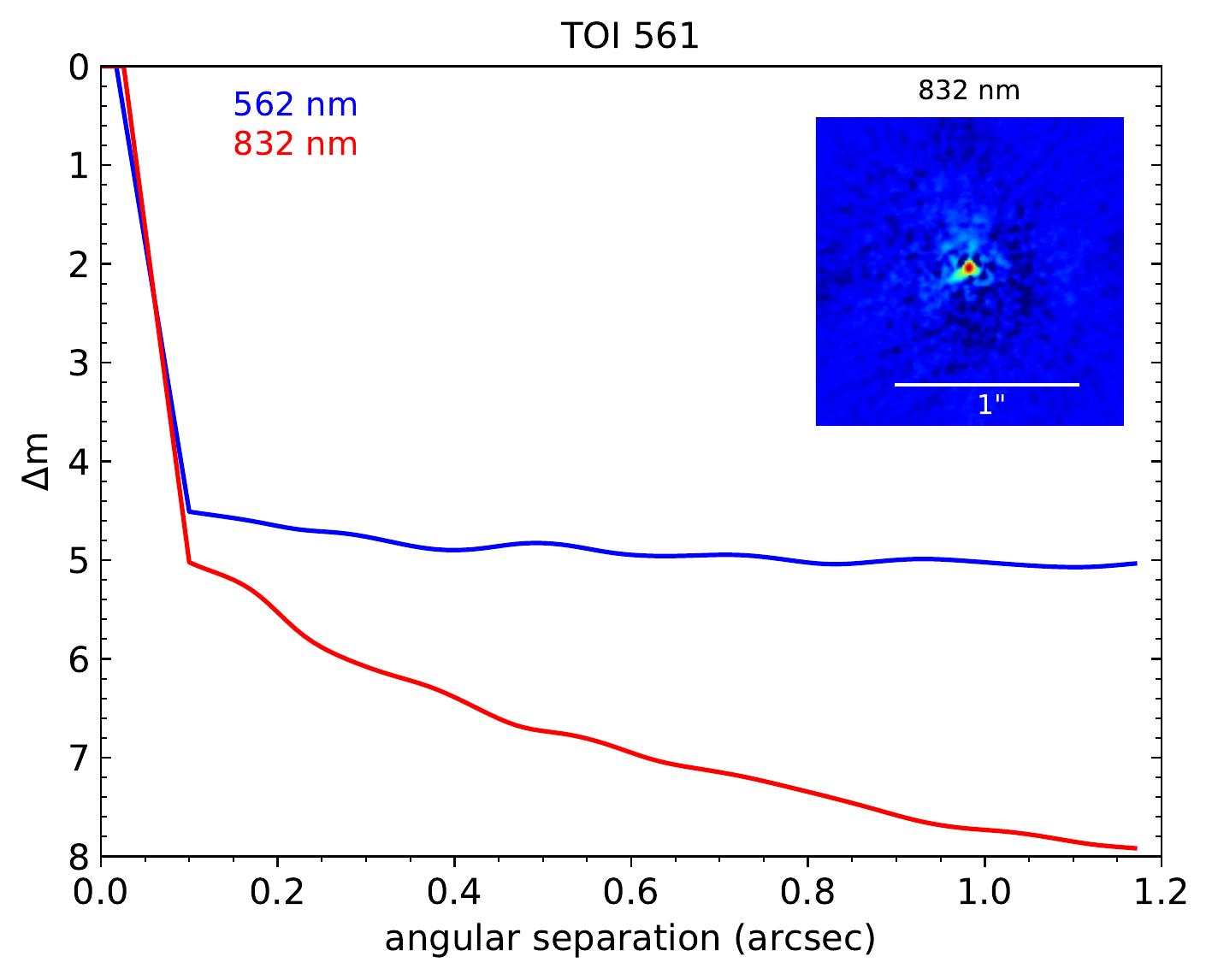}
    \caption{No nearby stars were detected in 4 independent imaging campaigns.  Top left: Sensitivity to background stars of our Gemini-N/NIRI images in the Br$\gamma$ filter. The images were taken in good seeing conditions, and we reach a contrast of 7.5 magnitues fainter than the host star within 0.\arcsec5. {\it Inset:} Image of the central portion of the data, centered on the star. Top right: Same as top left, but with Palomar. Bottom left: sensitivity to background stars from SOAR speckle imaging observations (5-$\sigma$ upper limits), with an example image inset.  Bottom right: Same as bottom left, but with Gemini-S/Zorro in two passbands.
}
    \label{fig:imaging}
\end{figure*}

The sensitivity curves are shown in Figure \ref{fig:imaging} along with an inset image zoomed to the primary target showing no other companion stars.   Both the Gemini and Palomar data reach a $\Delta$mag $\approx 2$ at 0.15\arcsec\ with an ultimate sensitivity of 7.7 mag and 8.7 mag for the Gemini and Palomar imaging, respectively.  To within the limits and sensitivity of the data, no additional companions were detected.

\subsection{SOAR and Gemini-South}
We also searched for stellar companions were also searched for with speckle imaging on the 4.1-m Southern Astrophysical Research (SOAR) telescope \citep{Tokovinin2018} on 18 May 2019 UT.  The speckle observations complement the NIR AO as the I-band observations are similar to the \textit{TESS} bandpass. More details of the observations are available in \citet{Ziegler2020}. The observations have a sensitivity of $\sim 1$ mag at a resolution of 0.06\arcsec\ and an ultimate sensitivity of $\sim 7$ mag at a radius of 3\arcsec.  The 5$\sigma$ detection sensitivity and speckle auto-correlation functions from the observations are shown in Figure \ref{fig:imaging}. As with the NIR AO data, no nearby stars were detected within 3\arcsec\ of TOI-561 in the SOAR observations.

High-resolution speckle interferometric images of TOI-561 were obtained on 15 March 2020 UT using the Zorro\footnote{https://www.gemini.edu/sciops/instruments/alopeke-zorro/} instrument mounted on the 8-meter Gemini South telescope located on the summit of Cerro Pachon in Chile. Zorro simultaneously observes in two bands, i.e., 832/40 nm and 562/54 nm, obtaining diffraction limited images with inner working angles 0.\arcsec017 and 0.\arcsec026, respectively. Our data set consisted of 5 minutes of total integration time taken as sets of $1000\times0.06$ second images. All the images were combined and subjected to Fourier analysis leading to the production of final data products including speckle reconstructed imagery \citep[see][]{Howell2011}. Figure \ref{fig:imaging} shows the 5$\sigma$ contrast curves in both filters for the Zorro observation and includes an inset showing the 832 nm reconstructed image. The speckle imaging results reveal TOI-561 to be a single star to contrast limits of 5 to 8 magnitudes, ruling out main sequence stars brighter than late M as possible companions to TOI-561 within the spatial limits of $\sim$2 to 103 au (at $d=86$ pc).

\section{Radial Velocities \label{sec:rvs}}
We obtained \nrvs\ high-resolution spectra with the W. M. Keck Observatory HIRES instrument on Maunakea, Hawaii between May 2019 and \lastobs\ 2020, at a cadence of one to two RVs per night.  We followed the standard observing and data reduction procedures of the California Planet Search \citep[CPS,][]{Howard2010_CPS}.  We obtained spectra with the C2 decker, which has dimensions of $14\arcsec \times 0\arcsec.86$ and spectral resolution R$\approx$60,000 at 500 nm.  We only observed when the target was at least $25^\circ$ from the moon.  At $V=10.2$, the star was always at least 8 magnitudes brighter than the moon-illuminated background sky.

\begin{figure}
    \centering
    \includegraphics[width=0.5\textwidth]{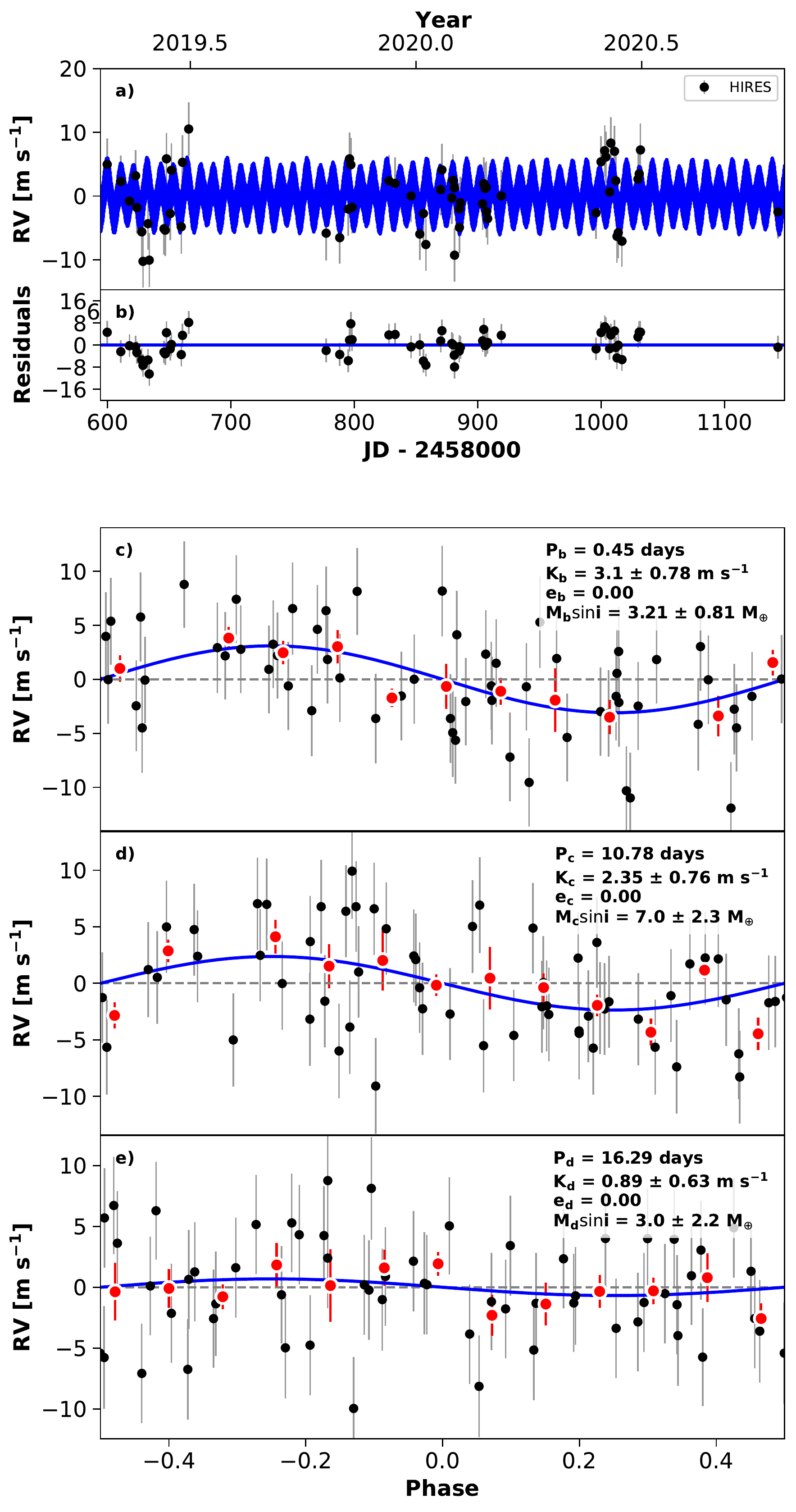}
    \caption{Radial velocities of TOI-561, based on observations from Keck-HIRES (black points).  Error bars are 1$\sigma$ confidence intervals, and the blue line is the best-fit model.  The top panel is the full RV time series and residuals.  Subsequent panels are the RVs components of planets b, c, and d, phase-folded to the orbital period of each planet (the model RV components from the other planets are subtracted from each panel).  The red points are RVs binned in phases of 0.1.
}
    \label{fig:rvs}
\end{figure}

We placed a warm cell of molecular iodine gas in the light path as a simultaneous wavelength calibration source for all RV spectra \citep{Marcy1992}.  We obtained a template spectrum by observing the star without the iodine cell.  We observed rapidly-rotating B stars, with the iodine cell in the light path, immediately before and after the template to model the PSF of the HIRES spectrograph.  Each RV spectrum was reproduced with a combination of the deconvolved template spectrum and a laboratory iodine atlas spectrum convolved with the HIRES PSF of the observation (which we empirically determined).  The RVs are listed in Table \ref{tab:rvs} and displayed in Figure \ref{fig:rvs}.  Before fitting for any planets, the RVs had an RMS of 5.0\,\ms, and the median individual RV error (before applying jitter) was 1.4\,\ms. 

\begin{deluxetable}{cccc}
\tablecaption{Radial Velocities \label{tab:rvs}}
\tablehead{\colhead{Time} & \colhead{RV} & \colhead{RV unc.} & \colhead{S val}\\
\colhead{(BJD)} & \colhead{(\ms)} & \colhead{(\ms)} & \colhead{} } 

\startdata
2458599.74193 & 3.9 & 1.3 & 0.148\\
2458610.76476 & 1.3 & 1.4 & 0.147\\
2458617.75866 & -1.9 & 1.4 & 0.142\\
2458622.74736 & 2.3 & 1.2 & 0.146\\
2458623.75550 & 2.9 & 1.4 & 0.146\\
2458627.75794 & -6.5 & 1.6 & 0.151\\
\enddata


\tablecomments{The first few lines are shown for form and content.  The full machine-readable table is available in the online version.}


\end{deluxetable}

When fitting RVs, the mass (and density) determinations of small planets can sensitively depend on the choice of model, particularly the number of planets included and their orbital periods and the use or non-use of correlated noise models.  For example, in the Kepler-10 system, the measured mass of Kepler-10 c ranged from 7 to 17 \mearth, based on the choice of model \citep{Dumusque2014, Weiss2016, Rajpaul2017}.  To test the robustness of our mass and density determinations, we applied several different models to the RVs of TOI-561.

\subsection{Three-Planet Keplerian Models}
We modeled the RVs with the publicly available \texttt{python} package \texttt{radvel} \citep{Fulton2018_radvel}.  We used the default basis of a five-parameter Keplerian orbit, in which the RV component of each planet is described by its orbital period ($P$), time of conjunction ($T_c$), eccentricity ($e$), argument of periastron passage ($\omega$), and RV semi-amplitude ($K$).  Because the orbital ephemerides from \textit{TESS} are more precise than what we can constrain with \nrvs\ RVs, we fixed $P$ and $T_c$ for each transiting planet at the best-fit values from \TESS\ photometry plus additional ground-based photometry from the \TESS\ Science Group 1.\footnote{We also tried allowing the periods and transit ephemerides to vary but with priors from the best-fit ephemeris, without substantial difference to the results.}  The eccentricities of all three planets are expected to be small for dynamical reasons.  At $P < 1$ day, the USP is almost certainly tidally circularized.  The other planets have compact orbits ($P_d/P_c\approx1.5$), such that large eccentricities would result in orbit crossings, and even modest eccentricities would likely result in Lagrange instability \citep{Deck2013}.  Furthermore, the majority of small exoplanets in compact configurations have low eccentricities \citep{VanEylen2019,Mills2019}.  For these reasons, and also because modeling eccentricities introduces two free parameters per planet, we only explored circular fits for all three transiting planets.

Thus, of the five Keplerian parameters that describe each transiting planet, only the semi-amplitude ($K$) was allowed to vary, along two global terms: an RV zeropoint offset ($\gamma$) and an RV jitter ($\sigma_j$), which is added to the individually determined RV errors in quadrature to account for non-Gaussian, correlated noise in the RVs from stellar processes and instrumental systematics.  Our full likelihood model was:
\begin{equation}
-2\mathrm{ln}\mathcal{L} = \sum_i \frac{(x_{\mathrm{meas},i} - x_{\mathrm{mod},i})^2}{\sigma_i^{\prime 2}} + \sum_i \rm{ln}(2\pi{\sigma^{\prime 2}_i})
\end{equation}

where
\begin{equation}
    \sigma^{\prime 2}_i = \sigma^2_i + \sigma^2_j
\end{equation}
is the quadrature sum of the internal RV error and the jitter.

We optimized the likelihood function with the Powell method \citep{Powell1964} and used a Markov-Chain Monte Carlo (MCMC) analysis\footnote{based on \texttt{emcee}, \citet{Foreman-Mackey2013}} to determine parameter uncertainties.  We explored the optimization of several models.  In Model A, we did not enforce any priors on planet semi-amplitudes (thus allowing values of $K$, and hence planet mass, to be negative).  Although negative planet masses are unphysical, their consideration offsets the bias toward high planet masses that occurs when planet masses are forced to be positive \citep{Weiss2014}.  The best-fit values with Model A were $K_b=\KbA\,\ms$, $K_c=\KcA\,\ms$, and $K_d=\KdA\,\ms$.  The RMS of the RV residuals was 4.2\,\ms.

In Model B, we restricted $K > 0$.  The advantages of restricting planet masses to be larger than zero are (1) the planet masses are physically motivated, and (2) the residuals are more likely to be useful in searching for additional planets.  Model B yielded $K_b=\KbB\,\ms$, $K_c=\KcB\,\ms$, and $K_d=\KdB\,\ms$.  The RMS of the RV residuals was 4.2\,\ms.

Model C was the same as Model B, except we allowed a linear trend in the RVs, $\dot{\gamma}$, which could be caused by acceleration from a long-period companion.  However, the RVs do not strongly favor a trend: $\dot{\gamma} = \gdot\,\rm{m s^{-1} day^{-1}}$, and the best-fit $K$ values changed by less than 1$\sigma$ with the inclusion of a trend: $K_b=\KbC\,\ms$, $K_c=\KcC\,\ms$, and $K_d=\KdC\,\ms$.  The RMS of the RV residuals was 4.0\,\ms.

In Model D, we considered the hypothesis that planet d has half of the presumed orbital period, which was possible given the gap in the photometry.  This model is the same as Model B, except $P_d=8$ days.  This model did not affect the amplitudes of planets b or c, but resulted in $K_d < 2.04\,\ms$ (2$\sigma$ confidence).  The RMS of the RV residuals was 4.1\,\ms.

In Models A-C, the choice of model makes very little impact on the best-fitting RV semi-amplitudes for each planet, and hence our planet mass determinations are robust with respect to our choice of model.  For the rest of this paper, we consider Model B as our default model unless stated otherwise.  The fitted and fixed parameters of Model B are provided in Table \ref{tab:planet}.

\begin{figure}
    \centering
    \includegraphics[width=\columnwidth]{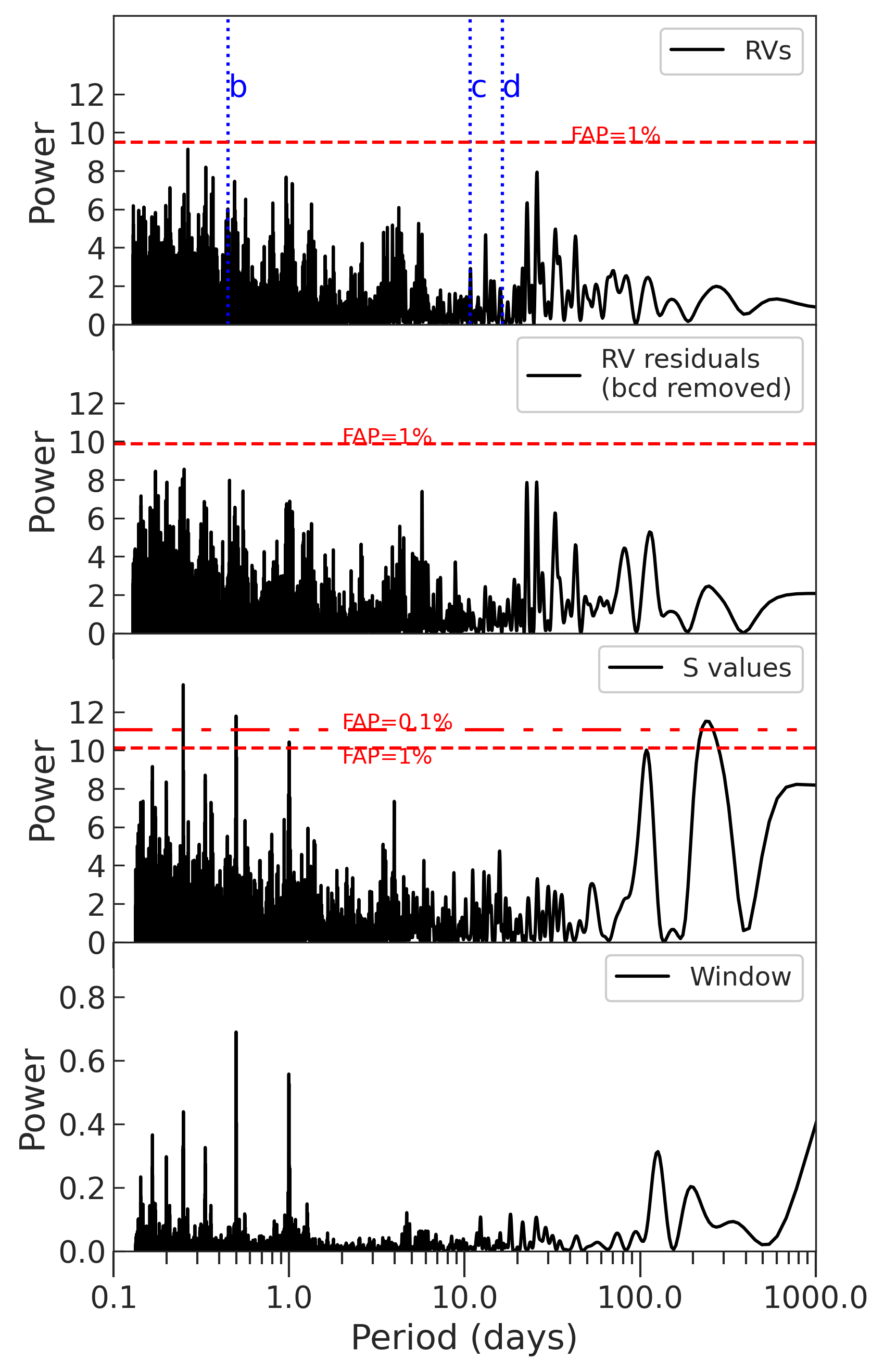}
    \caption{Lomb-Scargle periodograms of (a) the RVs, (b) the RVs after removing the best-fit [Model B] RV signatures from planets b, c, and d, (c) the Mt. Wilson S-value stellar activity indicator time series, and (d) the window function.  False alarm probabilities (FAPs) are computed based on a bootstrap resampling of each time series.  The RVs and RV residuals do not have any peaks that cross the 1\% FAP.  The S-values have significant (FAP $>1\%$) peaks at 100 days, 230 days, and various short periods that are likely aliases of the longer-period signals caused by the window function.}
    \label{fig:pergrams}
\end{figure}

\subsection{Correlated Noise Analysis}
The simple Keplerian model fit to observed RVs of TOI-561 displayed an RMS of 4.2\,\ms\ (Figure \ref{fig:rvs}), which is uncharacteristically high for precision RVs of such a bright star \citep{Howard2016}.  A Lomb-Scargle periodogram of the RV residuals reveals several peaks that might correspond to correlated noise and/or additional planets in the system.  However, none of the peaks were significant at the 1\% false alarm probability (FAP) level, which we determined with bootstrap resampling (Figure \ref{fig:pergrams}).

Nonetheless, we considered several correlated-noise models in an attempt to model and remove a putative red noise component responsible for the high RMS of the RV residuals.  In Model E, we employed Gaussian process regression (GP), which has been previously applied in analysing RVs of many exoplanets \citep[e.g. ][]{Haywood2014,Grunblatt2015}. For details of the GP model, see \citet{Dai2017}.  In principle, the light curve and RVs are both affected by stellar activity rotating in and out of view of the observer \citep[e.g.,][]{Aigrain2012}.  In an attempt to model the correlated noise in the RVs, we trained the various ``hyperparameters'' of our GP quasi-periodic kernel using the out-of-transit {\it TESS} PDCSAP light curve.  We did not detect the stellar rotation period in the {\it TESS} PDCSAP or SAP light curves, possibly because the star is very inactive (log$R^\prime_{\mathrm{HK}} = -5.1$), or the expected rotation period of the star is longer than the photometric baseline.  Our MCMC fit to the lightcurve produced broad posteriors on the rotation period and the other hyperparameters. Therefore, we did not impose Gaussian priors on the hyperparameters in our GP analysis of the RV data, although we did limit the stellar rotation period $<$500 days for quicker convergence. We  used {\sc emcee} to constrain the posterior distribution of the GP  hyperparameters simultaneously with the orbital parameters of the three planets. We detect the RV signal of planet b: $K_b = 2.9 \pm 0.7$ m$~$s$^{-1}$ and planet c: $K_b = 1.7 \pm 1.0$ and an upper limit for planet d: $K_d < 1.6$ m$~$s$^{-1}$ (95\% confidence), all of which are within 1$\sigma$ of the planet semi-amplitudes we determined without the correlated noise component of the model.

We also attempted to train a correlated noise model on our spectroscopically determined Mt. Wilson S-values.  A Lomb-Scargle periodogram of the S-values shows two significant peaks: one at 100 days, and the other at 230 days, both of which cross the 1\% FAP threshold (which we computed with bootstrap resampling, Figure \ref{fig:pergrams}).  (The forest of regular peaks with $P < 1$ day are likely aliases produced by our window function, which had poor sampling below 1 day).  Note that these periods differ from the most prominent peaks in the RV residuals, which are a doublet near 25 days.  Futhermore, the S-values are not sampled as frequently as the lightcurve, but they are sampled simultaneously with the RVs, and they are a direct indicator of the chromospheric magnetic activity during the observations.  We tried a GP with a quasi-periodic kernel trained on the S-values (Model F), but this did not produce significant changes in the semi-amplitudes of the planets or the RMS of the RV residuals, possibly because the S-values had small variability or were sparsely sampled.  We also tried a model in which we decorrelated the RVs with respect to the S-values (Model G), which did not reduce the RMS (and thus did not remove the correlated noise).

None of our attempts to model correlated noise in the RVs changed the amplitudes of the planets or reduced the RMS of the RV residuals, and so we prefer the simpler Keplerian models (without correlated noise) described above.  Perhaps TOI-561 is too inactive for models trained on stellar activity to be effective, given the current quality of the data.  For comparison, Kepler-10, another system with time-correlated RV residuals, has log$R^\prime_\mathrm{HK} = -4.89$, which is more active than TOI-561 log$R^\prime_\mathrm{HK} = -5.1$.  The use of Gaussian processes  affected the mass determination of Kepler-10 c, lowering it from 14\,\mearth\ (no GP) to 7\,\mearth\ (with GP).  Perhaps there is a minimum stellar activity for which attempts to decorrelate the stellar activity signal can be successful, given RVs with precision of 2\,\ms.

Nonetheless, there are substantial correlated residuals in the RVs of TOI-561 which are uncharacteristic of the HIRES instrument performance \citep[typically 2\,\ms\ for $V < 11$,][]{Howard2016}.  The residual RVs of TOI-561 are not well-explained by any of our models of stellar activity, and so perhaps additional planets contribute to the RV residuals.  More RVs are needed to identify the orbital periods of any such planets and model their Keplerian signals.

\section{Planet Masses \& Densities \label{sec:masses}}
Each $K$ value can be converted to the planet's minimum mass, $\mplsini$, but because all three planets transit, actual masses (rather than minimum masses) can be calculated.  Assuming Model B, we find $M_b=\mb\,\mearth$, $M_c=\mc\,\mearth$, and $M_d=\md\,\mearth$.  Furthermore, since the planets transit, their radii are calculated from the planet-to-star radius ratios and known stellar radius: $R_b=\rb\,\rearth$, $R_c=\rc\,\rearth$, and $R_d=\rd\,\rearth$.  The bulk densities of the planets are $\rho_b=\rhob\,\gcc$, $\rho_c=\rhoc\,\gcc$, and $\rho_d=\rhod\,\gcc$.  The derived physical and orbital properties of the planets are summarized in Table \ref{tab:planet}.

\begin{deluxetable}{lc}
\tablecaption{Planet Parameters}\label{tab:planet}
\tablehead{\colhead{Parameter} & 
  \colhead{Median $\pm$ 1$\sigma$}
}
\startdata
\multicolumn{2}{c}{Planet b} \\
\hline
Orbital Period, $P_b$ (days) & $0.446573^{+0.000032}_{-0.000021}$\\
Mid-Transit Time, $T_c$ (BJD)  & $2458517.4973\pm0.0018$\\
Radius Ratio, $R_p/R_\star$ & $0.016\pm0.001$ \\
Impact Parameter, b & $0.3 \pm 0.2$ \\
Duration, $T_{14}$ (hours) & $1.42 \pm 0.10$ \\
Orbital Eccentricity, e & 0 (fixed) \\
RV semi-amplitude, $K_{b}$ (\ms) & \KbB \\
Semi-major axis, $a_b$ (AU) & $0.01064\pm 0.00013$  \\
Radius, $R_b$ (\rearth) & \rb \\
Mass, $M_b$ (\mearth) & \mb \\
Density, $\rho_b$ (\gcc) & \rhob \\
Equilibrium temperature, $T_{\rm{eq,b}}$ (K) & $2480\pm200$\\
\hline
\multicolumn{2}{c}{Planet c} \\
\hline
Orbital Period, $P_c$ (days) & $10.77892\pm0.00015$\\
Mid-Transit Time, $T_c$ (BJD) & $2458527.05825\pm0.00053$\\
Radius Ratio, $R_p/R_\star$ & $0.032\pm0.001$\\
Impact Parameter, b & $0.2 \pm 0.2$ \\
Duration $T_{14}$ (hours) & $4.04 \pm 0.26$ \\
Orbital Eccentricity, e & 0 (fixed) \\
RV semi-amplitude, $K_{c}$ (\ms) & \KcB \\
Semi-major axis, $a_c$ (AU) & $0.0888\pm 0.0011$\\
Radius, $R_c$ (\rearth) & \rc \\
Mass, $M_c$ (\mearth)  & \mc \\
Density, $\rho_c$ (\gcc) & \rhoc \\
Equilibrium temperature, $T_{\rm{eq,c}}$ (K) & $860\pm70$\\
\hline
\multicolumn{2}{c}{Planet d} \\
\hline
Orbital Period, $P_d^{\dagger}$ (days) & $16.287 \pm 0.005$ \\
Mid-Transit Time, $T_c$ (BJD) & $2458521.8828 \pm 0.0035$ \\
Radius Ratio, $R_p/R_\star$ & $0.0256 \pm 0.0016$\\
Impact Parameter, b & $0.1 \pm 0.1$\\
Duration, $T_{14}$ (hours) & $4.45 \pm 0.46$\\
Orbital Eccentricity, e & $0$ (fixed)\\
RV semi-amplitude, $K_d$ (\ms) & \KdB \\
Semi-major axis, $a_d^{\dagger}$ (AU) & $0.1174\pm 0.0015$\\
Radius, $R_d$ (\rearth) & \rd \\
Mass, $M_d^{\dagger}$ (\mearth)  &  \md \\
Density, $\rho_d^{\dagger}$ (\gcc)  & \rhod \\
Equilibrium temperature, $T_{\rm{eq,d}}^{\dagger}$ (K)  & $750\pm60$\\
\hline
\multicolumn{2}{c}{Other} \\
\hline
RV Zeropoint, $\gamma$ (\ms) & $-0.8291$ \\
RV Jitter, $\sigma_{\rm{j}}$ (\ms) & $4.09^{+0.49}_{-0.42}$ \\
\enddata
\tablenotetext{\dagger}{The orbital period of planet d was incorrectly identified as $P=16.37$ days in the original SPOC pipeline because of a partially masked transit.  An alias of the orbital period of planet d, $P_d=8$ days, is also consistent with the data.  Parameters marked with a dagger would be affected by an incorrect assumption of the orbital period of planet d.  See \S\ref{sec:discussion} for assumptions regarding planet equilibrium temperatures.}
\end{deluxetable}

\begin{figure}
    \centering
    \includegraphics[width=\columnwidth]{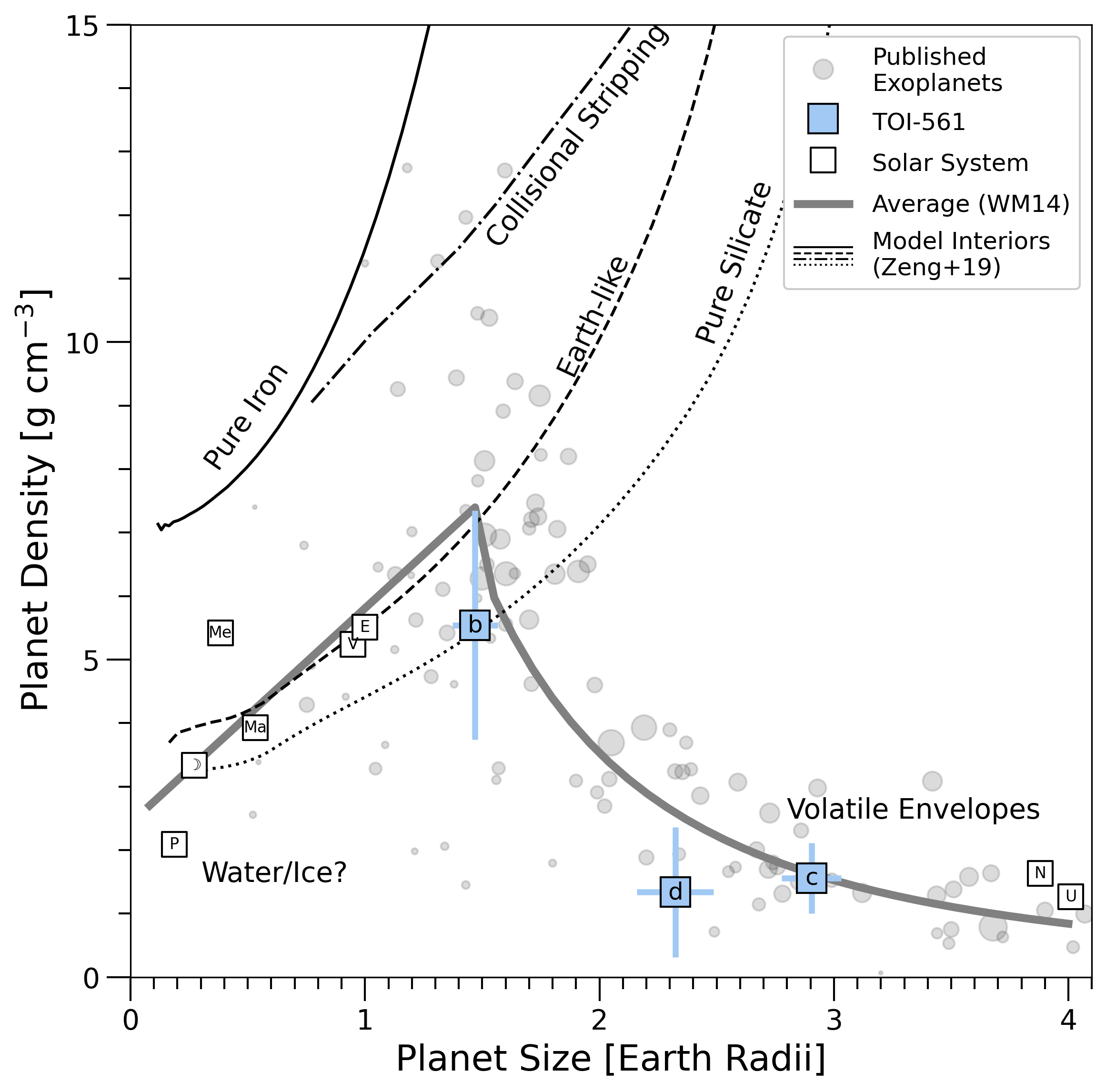}
    \caption{Planet bulk density vs. planet radius for small planets with measured radii ($\rpl < 4\,\rearth$, $\sigma(\rpl)/\rpl < 0.2$) and masses ($\sigma(\mpl) < 2\,\mearth$), based on results from the NASA Exoplanet Archive \citep[queried 2020 June 24][gray points]{Akeson2013}.  The point size is scaled to $\rhopl/\sigma(\rhopl)$.  The \citet{Weiss2014} average mass-radius relationship, several model composition curves from \citet{Zeng2019}, and the theoretial maximum density from collisional stripping \citep{Marcus2010} are shown.  The solar system planets and Earth's moon are provided for context.  The TOI-561 planets are consistent with or slightly less dense than typical planets of their sizes.  TOI-561 b is 1$\sigma$ less dense than an Earth-composition planet.}
    \label{fig:mr}
\end{figure}

The masses and densities of the TOI-561 planets are shown in comparison to the masses and densities of other sub-Neptune sized planets in Figure \ref{fig:mr}.  The other planet masses and densities come from the NASA Exoplanet Archive, from which we included only those with $\sigma(\mpl) < 2\,\mearth$.

\subsection{TOI-561 b}
The USP TOI-561 b has a typical mass and density for its size.  At 1.5\,\rearth\ and \rhob\,\gcc, it is 1$\sigma$ below the peak of the density-radius diagram identified in \citet{Weiss2014}, consistent with a rocky composition that is either Earth-like or iron-poor.  Nearly all USPs are smaller than 2\,\rearth\ and are expected to have rocky compositions, given their small sizes and extreme stellar irradiation \citep{Sanchis-Ojeda2015}, and TOI-561 b is consistent with this expectation.   In a homogeneous analysis of USPs with masses determined from RVs, \citet{Dai2019} found that most USPs with $<10\,\mearth$ are consistent with having Earth-like compositions, whereas the few USPs with $>10\,\mearth$ likely have H/He envelopes. TOI-561 b (\mb\,\mearth) is consistent with the rocky group of that study.

The minimum density of a USP can be determined from its orbital period and the requirement that it orbits outside the Roche limiting distance \citep{Rappaport2013}.  We investigated the minimum density of TOI-561 b, with the hope that it would provide additional constraints on the mass of the planet.  Using the approximation from \citet{Sanchis-Ojeda2014}, we find that the minimum density of the USP is

\begin{equation}
\rhopl \mathrm{[\gcc]} \ge (11.3 \mathrm{hr}/P_{\mathrm{orb}})^2 = 1.15 \gcc
\end{equation}
for TOI-561 b, which is below our measured density and corresponds to a minimum mass of 0.64\,\mearth.  Such a low mass is ruled out by the data at nearly 3$\sigma$ confidence.  Thus, TOI-561 b is not close enough to its star for the Roche stability criterion to provide additional information about the density.

An interpretation of the USP composition that involves a H/He layer is disfavored by physical models.  At a sufficiently low stellar irradiation level, the low density of TOI-561 b might have been consistent with a composition of an Earth-like core overlaid with a thin H/He envelope.  However, at $P=0.44$ days (with equilibrium temperature $>2000\,K$) and a mass of $\sim 3\,\mearth$, the USP is too irradiated and too low-mass to hold onto a H/He envelope \citep{Lopez2017}.

\subsection{TOI-561 c \& d}
At \rpl > 1.5\,\rearth\ and with low densities, TOI-561 c and d have substantial gaseous envelopes by volume, although the gas envelopes likely only constitute $\sim1\%$  of the planet masses \citep{Lopez2014}.  TOI-561 c has a radius and mass consistent with the \citet{Weiss2014} empirical mass-radius relationship.

The ambiguity of the orbital period for planet d poses a challenge to accurate mass determination.  Our RVs are consistent with a non-detection of planet d at both the $P_d=16$ and $P_d=8$ day orbits.  Assuming $P_d=16$ days, the RVs provide an upper limit of $K_d < 2.1\,\ms$, which corresponds to $\mpl < 7.0\,\mearth$ (2$\sigma$ confidence, see \S\ref{sec:rvs}).  Assuming $P_d=8$ days, the RVs provide an upper limit of $K < 2.0\,\ms$, which corresponds to $\mpl < 5.6\,\mearth$ (2$\sigma$ confidence).  In either scenario, the mass of TOI-561 d is approximately 2$\sigma$ below the \citet{Weiss2014} mass-radius relationship and is too low to be consistent with a rocky composition, given the planet's radius.  However, if planet d is actually the transits of two distinct planets as suggested in \citet{Lacedelli2020}, the low mass presented here does not apply.

\subsection{Non-Detection of Outer Giant Planets}
With our 1.5 year baseline of RVs, we were able to place constraints on possible long-period giant planets.  In Model C, we found a 3$\sigma$ upper limit to an RV trend of $\dot{\gamma} < 0.02 \ms \mathrm{day}{-1}$.  This observational limit on the acceleration of the star can be converted to a limit on the mass and orbital distance of a possible perturber by setting it equal to the gravitational acceleration from a planet at distance $r$: $G M_\star \mpl r^{-2} = M_\star \dot{\gamma}$, where G is the gravitational constant and $r$ is the distance between the perturber and the star at the time the RVs were measured.  Assuming a circular orbit for the putative giant planet ($r=a$), we find
\begin{equation}
    \frac{\mplsini}{\mjup} (\frac{a}{5 \mathrm{AU}})^{-2} < 1.0
\end{equation}
(3$\sigma$ confidence).  Thus, our non-detection of an RV acceleration rules out a 1.0\,\mjup\ planet at 5 AU with 3$\sigma$ confidence (assuming sin$i\approx1$, $e=0$, and that we did not primarily sample the orbit while the planet was moving parallel to the sky plane).  In systems with compact configurations of transiting planets, giant planets often need to be approximately coplanar with the transiting planets in order for the inner planets to remain mutually continuously transiting \citep{Becker2017}, and so we do not expect non-transiting companions to have face-on orbits.  Thus, the non-detection of an RV trend rules out a variety of scenarios of a coplanar giant planet near the snow-line.  

\section{Discussion \label{sec:discussion}}
\subsection{Stability}
We investigated whether we could constrain the architecture of the TOI-561 system, and in particular the orbital period of planet d, with stability arguments.  To asses stability, we used \texttt{Spock}, a machine-learning based approach to inferring orbital stability \citep{Tamayo2020}.  Spock incorporates several analytic indicators (including MEGNO, AMD, and Mutual Hill radius) with an N-body integration of $10^4$ orbits to compute a probabilistic assessment of the system stability.  We tested the architectures of TOI-561 with planet d at $P_d=16$ days (Model B) and at $P_d=8$ days (Model D).  Since stability depends sensitively on the eccentricities and inclinations of planets in compact configurations, we varied the eccentricities, arguments of periastron passage, and inclinations.  The eccentricities and inclinations in multiplanet systems are well-fit with Rayleigh distributions with characteristic values $\sigma_e = 0.035$ and $\sigma_i =2.45^\circ$ \citep{Mills2019}, and so we sampled 1000 trials from these distributions (sampling $\omega$ uniformly on [0,$\pi$]).  For Model B, the probability of stabiltiy is $95\pm3\%$, whereas for Model D, the probability of stability is $84\pm 30\%$.  Thus, although more configurations of Model B are stable, we cannot rule out Model D based on a stability argument.

\subsection{Formation and Evolution}
TOI-561 is the second transiting multiplanet system discovered around a galactic thick-disk star \citep[Kepler-444,][is the first]{campante15}, and is the first galactic thick disk star with a USP.  The iron-poor, alpha-enhanced stellar abundances observed today are likely representative of the nebular environment in which the planets formed.  Thus, TOI-561 provides an opportunity to study the outcome of planet formation in an environment that is chemically distinct from most of the planetary systems known to date.  Furthermore, its membership in the galactic thick disk indicates old age, making the rocky planet TOI-561 b one of the oldest rocky planets known. 

The confirmed old age of the system is relevant for dynamical studies of the planets.  For instance, the formation of USPs is still poorly understood.  The majority of USPs are the only detected transiting planet in their systems \citep{Sanchis-Ojeda2014}.  This is partially due to a bias of geometry (as planet detection probability scales with $R\star/a$).  However, \citet{Dai2018} found that the mutual inclinations in multiplanet systems with USPs are significantly larger than in the multiplanet systems without USPs, suggesting that systems with USPs have had more dynamically ``hot'' histories.  One mechanism for generating large mutual inclinations in systems with USPs is if the host star is oblate and misaligned with respect to the planets \citep{Li2020}.  

Furthermore, dynamical interactions in a multiplanet system can move short-period planets to ultra-short periods in a manner that may excite large eccentricities.  \citet{Petrovich2019} proposed a mechanism of secular chaos in which the innermost planet is kicked to a high-eccentricity orbit, which is then circularized.  In contrast, \citet{Pu2019} proposed a scheme in which the innermost planet's neighbors consistently force it to a low-eccentricity orbit, which results in inward migration and eventual circularization.

In a study of the Gaia-DR2 kinematics of USP host stars in the \Kepler\ field, \citet{Hamer2020} found that their motions are similar to those of matched field stars (rather than young stars).  The broad range of ages of USPs suggests that USPs do not undergo rapid tidal inspiral during the host star's main sequence lifetime.  The existence of USP around an 10 Gyr old star is consistent with this finding. 

\subsection{Predicted Transit Timing Variations}
Because planets c and d have an orbital period ratio of $\sim1.5$ or less, the planets likely perturb each other's orbits, producing transit timing variations (TTVs).  The \textit{TESS} sector 8 baseline was too short to detect TTVs (only two transits of each planet were detected).  The approximate amplitude of the TTV signal can be computed analytically using the expressions from \citet{Lithwick12b}:

\begin{equation}
\label{eqn:ttv}
    |V| \approx P \mu^\prime/\Delta + \mathcal{O}(e) / \Delta
\end{equation}
where $P$ is the orbital period of the inner planet, $\mu^\prime$ is the mass of the perturbing, outer planet (in units of stellar mass), $\mathcal{O}(e)$ is a first-order dependency on the free eccentricities of the planets, and $\Delta$ is the non-dimensional distance from mean motion resonance:
\begin{equation}
    \Delta = \frac{ P^\prime }{P} \frac{j-1}{j} - 1
\end{equation}
where $P$ and $P^\prime$ are the inner and outer orbital periods and $j$ is an integer.  Equation \ref{eqn:ttv} can be used to approximate the TTV amplitude of the outer planet by setting $P \longrightarrow P^\prime$ and $\mu^\prime \longrightarrow \mu$ \citep[for a thorough derivation and caveats, see][]{Lithwick12b}.

Using the values for the planet masses determined in \S\ref{sec:masses} and assuming circular orbits, we find that the TTV amplitude of planet c ($P=10.78$ days) is about 6 minutes, whereas the TTV amplitude of planet d is either 30 minutes (if $P_d = 16$ days; $j=3$) or 15 minutes (if $P_d = 8.14$ days, $j=4$).

\subsection{Atmospheric Probing Metrics}
Using the system parameters tabulated in this paper, we calculated the Transmission and Emission Spectroscopy Metrics (TSM and ESM, respectively) of \cite{kempton:2018} to determine whether these newly-characterized planets are compelling targets for future atmospheric or surface characterization via transit or eclipse spectroscopy.  Assuming a flat prior on the planets' Bond albedos from 0 to 0.4 and on their efficiency of day-to-night energy recirculation from 1/4 to 2/3, and propagating the uncertainties on all relevant parameters, we calculate the planets' equilibrium temperatures to be $2480\pm200$ K, $860\pm70$K, and $750\pm60$ K for planets b, c, and d, respectively.  Because of the relatively shallow transit depths ($\lesssim 1000$~ppm for all three planets), space-based spectroscopy is likely to be the only feasible avenue for such studies.
In contrast to some recent reports of these quantities for other newly-discovered \textit{TESS} planets, here we propagate all parameter uncertainties in order to report how well the TSM and ESM are constrained, as well as how promising the median values are; this is especially essential when planetary properties are not yet measured to high precision. We report the transmission and emission metrics and their uncertainties in Table~\ref{tab:atmo}.

Our analysis shows that TOI-561~b is among the best \textit{TESS} targets discovered to date for thermal emission measurements \citep[cf.\ Table 3 of][]{astudillo:2020}.    With $\mathrm{ESM} =7.1 \pm 1.1$, planet b is clearly a promising target for observations of its  secondary eclipse and/or its full-orbit phase curves, as has previously been done for other irradiated terrestrial planets  planets such as 55~Cnc~e \citep{demory:2012,demory:2016b} and LHS~3844b \citep{kreidberg:2019}. The uncertainty on planet b's ESM is dominated by the uncertainty on its transit depth, but regardless the planet has a reliably high metric in this category. Because of their cooler temperatures, the lower ESM values for planets b and c mark them as less attractive targets for secondary eclipse studies.
   
As for transit spectroscopy, the TSM values for the sub-Neptunes TOI-561c and~d listed in Table~\ref{tab:atmo} ($93^{+55}_{-37}$ and $82^{+109}_{-33}$, respectively) indicate that they {\em may} be particularly amenable to transmission studies. However, because the TSM scales inversely with planetary surface gravity this result depends on determining more precise values of these planets' masses. These planets clearly warrant additional precise RV followup: if the expectation values of their TSMs do not change as the uncertainties shrink, these two planets would be among the top 20 confirmed warm Neptunes for transmission spectroscopy \citep[cf.\ Table 11 of][]{guo:2020}.  Due to the small size of the highly irradiated planet b, and because it is unlikely to have retained much of an atmosphere, it is not an appealing target for transmission measurements.

Better ephemerides for planets c and d are necessary in preparation for atmospheric studies.  The alias of the orbit of planet d should be resolved prior to interpretation of the planetary atmospheres, since the factor of two change in orbital period produces a factor of $\sim1.3$ change in the equilibrium temperature.  Also, planet d may have significant TTVs with amplitudes of $\sim30$ minutes (assuming the orbital period is 16 days).

\begin{deluxetable}{lccl}
\tablecaption{Atmospheric Prospects}\label{tab:atmo}
\tablehead{\colhead{Planet} & 
\colhead{TSM} & 
\colhead{ESM} & 
\colhead{Notes}
}
\startdata
b & $11^{+73}_{-4}$&  $8.7 \pm 1.1$ &  Good eclipse target \\
c & $93^{+55}_{-37}$ &  $4.87 \pm 0.39$  &  Promising transmission target \\
d$^\dagger$ & $82^{+109}_{-33}$  &   $2.15 \pm 0.29$  &  Promising transmission target \\
\enddata
\tablenotetext{\dagger}{Assuming $P_d = 16$ days.  If $P_d = 8$ days then TSM$_d = 103^{+137}_{-43}$ (68\% confidence interval).}
\end{deluxetable}

\section{Conclusion \label{sec:conclusion}}
TOI-561 is a system with multiple transiting planets identified by the NASA \textit{TESS} spacecraft.  In this paper, we have confirmed two of the planets, including a rocky ultra-short period planet, with ground-based follow-up, and also characterized the properties of the planet and the host star.  We found:
\begin{enumerate}
    \item TOI-561 is a metal-poor, alpha-enhanced member of the galactic thick disk ($\feh=-0.4, \alpha=0.2$).  It is one of the oldest planetary systems yet identified and one of the most metal-poor.  In both of these aspects it is an important benchmark in our understanding of planet formation and evolution.
    \item We confirm the planets b (TOI-561.02, $P_b=0.45$ days, $R_b=\rb\,\rearth$) and c (TOI-561.01, $P_c=10.78$ days, $R_c=\rc\,\rearth$) with RVs, high-resolution imaging, and ground-based photometry.  We rule out a variety of astrophysical false positives for planet d (TOI-561.03, $P_d=16.29$ days, $R_d=\rd\,\rearth$) but note that the ephemeris is highly uncertain.
    \item With \nrvs\ RVs from Keck-HIRES, we determined the mass and density of the ultra-short period rocky planet TOI-561 b: $M_b = \mb\,\mearth$, $\rho_b = \rhob\,\gcc$.  Planet b has a below average density for its size (by 1$\sigma$), suggesting an iron-poor composition in the core.
    \item We also determined the mass and density of planet c ($M_c = \mc\,\mearth$, $\rho_c = \rhoc\,\gcc$) and an upper limit for the mass of planet d assuming $P_d=16$ days ($M_d = \md\,\mearth$).  The large radii and low masses of planets c and d are consistent with thick volatile envelopes overlaying rocky cores.
    \item The RVs from Keck-HIRES span 1.5 years and do not have a significant trend.  The non-detection of a trend rules out various scenarios of a giant planet near the ice line.
    \item Thanks to the bright host star, this multi-planet system is amenable to atmospheric follow-up with space-based telescopes.  Planet b is expected to be a good eclipse target, while planets c and d are promising targets for transmission spectroscopy.  Comparative atmospheric properties for the planets in this very metal-poor system would provide a unique test for planet formation scenarios.
\end{enumerate}

\smallskip
We thank the time assignment committees of the University of California, the California Institute of Technology, NASA, and the University of Hawaii for supporting the TESS-Keck Survey with observing time at the W. M. Keck Observatory and on the Automated Planet Finder.  

We thank NASA for funding associated with our NASA-Keck Key Strategic Mission Support project.  We gratefully acknowledge the efforts and dedication of the Keck Observatory staff for support of HIRES and remote observing.  We recognize and acknowledge the cultural role and reverence that the summit of Maunakea has within the indigenous Hawaiian community. We are deeply grateful to have the opportunity to conduct observations from this mountain.  

We thank Ken and Gloria Levy, who supported the construction of the Levy Spectrometer on the Automated Planet Finder. We thank the University of California and Google for supporting Lick Observatory and the UCO staff for their dedicated work scheduling and operating the telescopes of Lick Observatory.  This paper is based on data collected by the \textit{TESS} mission. Funding for the \textit{TESS} mission is provided by the NASA Explorer Program.  We acknowledge the use of public \textit{TESS} Alert data from pipelines at the \textit{TESS} Science Office and at the \textit{TESS} Science Processing Operations Center.  We thank David Latham for organizing the TESS community follow-up program, which brought together the widespread authorship and diversity of resources presented in this manuscript.

The work includes observations obtained at the international Gemini Observatory, a program of NSF's NOIRLab acquired through the Gemini Observatory Archive at NSF's NOIRLab, which is managed by the Association of Universities for Research in Astronomy (AURA) under a cooperative agreement with the National Science Foundation on behalf of the Gemini Observatory partnership: the National Science Foundation (United States), National Research Council (Canada), Agencia Nacional de Investigaci\'{o}n y Desarrollo (Chile), Ministerio de Ciencia, Tecnolog\'{i}a e Innovaci\'{o}n (Argentina), Minist\'{e}rio da Ci\^{e}ncia, Tecnologia, Inova\c{c}\~{o}es e Comunica\c{c}\~{o}es (Brazil), and Korea Astronomy and Space Science Institute (Republic of Korea). Data were collected under program GN-2019A-LP-101.  Observations in the paper made use of the High-Resolution Imaging instrument Zorro. Zorro was funded by the NASA Exoplanet Exploration Program and built at the NASA Ames Research Center by Steve B. Howell, Nic Scott, Elliott P. Horch, and Emmett Quigley. Zorro was mounted on the Gemini South telescope of the international Gemini Observatory.  Observations also made use of the NIRI instrument, which is mounted at Gemini North.  The Gemini North telescope is located within the Maunakea Science Reserve and adjacent to the summit of Maunakea. We are grateful for the privilege of observing the Universe from a place that is unique in both its astronomical quality and its cultural significance.  

This work makes use of observations from the LCOGT network.  This article is based on observations made with the MuSCAT2 instrument, developed by ABC, at Telescopio Carlos Sánchez operated on the island of Tenerife by the IAC in the Spanish Observatorio del Teide.  This work is partly supported by JSPS KAKENHI Grant Numbers JP17H04574, JP18H01265 and JP18H05439, and JST PRESTO Grant Number JPMJPR1775.  This work makes use of data collected under the NGTS project at the ESO La Silla Paranal Observatory.  The NGTS facility is operated by the consortium institutes with support from the UK Science and Technology Facilities Council (STFC)  projects ST/M001962/1 and  ST/S002642/1.

This work has made use of data from the European Space Agency (ESA) mission
{\it Gaia} (\url{https://www.cosmos.esa.int/gaia}), processed by the {\it Gaia}
Data Processing and Analysis Consortium (DPAC,
\url{https://www.cosmos.esa.int/web/gaia/dpac/consortium}). Funding for the DPAC
has been provided by national institutions, in particular the institutions
participating in the {\it Gaia} Multilateral Agreement.
JSJ acknowledges support by FONDECYT grant 1201371, and partial support from CONICYT project Basal AFB-170002.
JIV acknowledges support of CONICYT-PFCHA/Doctorado Nacional-21191829.

This research has made use of the Exoplanet Follow-up Observing Program (ExoFOP), which is operated by the California Institute of Technology, under contract with the National Aeronautics and Space Administration.  Resources supporting this work were provided by the NASA High-End Computing (HEC) Program through the NASA Advanced Supercomputing (NAS) Division at Ames Research Center for the production of the SPOC data products. 

L.M.W. is supported by the Beatrice Watson Parrent Fellowship and NASA ADAP Grant 80NSSC19K0597.  D.H. acknowledges support from the Alfred P. Sloan Foundation, the National Aeronautics and Space Administration (80NSSC18K1585, 80NSSC19K0379), and the National Science Foundation (AST-1717000).
E.A.P. acknowledges the support of the Alfred P. Sloan Foundation. 
C.D.D. acknowledges the support of the Hellman Family Faculty Fund, the Alfred P. Sloan Foundation, the David \& Lucile Packard Foundation, and the National Aeronautics and Space Administration via the \textit{TESS} Guest Investigator Program (80NSSC18K1583).  
I.J.M.C. acknowledges support from the NSF through grant AST-1824644.
Z.R.C. acknowledges support from the TESS Guest Investigator Program (80NSSC18K18584).
A.C. acknowledges support from the National Science Foundation through the Graduate Research Fellowship Program (DGE 1842402).
P.D. acknowledges support from a National Science Foundation Astronomy and Astrophysics Postdoctoral Fellowship under award AST-1903811. 
A.B. is supported by the NSF Graduate Research Fellowship, grant No. DGE 1745301.
R.A.R. is supported by the NSF Graduate Research Fellowship, grant No. DGE 1745301.
M.R.K. is supported by the NSF Graduate Research Fellowship, grant No. DGE 1339067.
J.N.W. thanks the Heising Simons Foundation for support.

\facilities{TESS, KeckI-HIRES, Gemini-North-NIRI, Gemini-South-Zorro, Palomar, SOAR, LCOGT, NGTS, El Sauce, PEST, MuSCAT2}
\software{Astropy \citep{Astropy_I,Astropy_II}, radvel \citep{Fulton2018_radvel}, emcee \citep{Foreman-Mackey2013}, Spectroscopy Made Easy \citep{Valenti:1996_sme,Piskunov:2017_sme_evolution}, SpecMatch Synth \citep{Petigura2015PhD}, SpecMatch-Emp  \citep{Yee2017}, isoclassify \citep{huber17}, AstroImageJ \citep{Collins:2017}, lightkurve \citep{Lightkurve2018}, spock \citep{Tamayo2020}, kiauhoku \citep{Claytor2020}}


\bibliography{main}{}
\bibliographystyle{aasjournal}
\allauthors
\end{document}